\documentclass[3p,twocolumn]{elsarticle}

\usepackage{amsmath}
\usepackage{subcaption}

\usepackage{amssymb}
\usepackage[figuresright]{rotating}
\usepackage{xcolor}
\usepackage{caption}
\usepackage{microtype}
\usepackage{diagbox}
\usepackage{cleveref}
\usepackage{csquotes}

\def\appendixname{}

\begin{document}

\begin{frontmatter}




\title{Confined Growth with slow surface kinetics: a Thin Film Model approach}%\\
%Morphology diagram and shape of a crystal growing on a substrate}


\author{Luca Gagliardi\corref{cor1}}
\ead{luca.gagliardi@univ-lyon1.fr}
\author{Olivier Pierre-Louis}
\ead{olivier.pierre-louis@univ-lyon1.fr}

\address{Institut Lumi\`ere Mati\`ere, UMR5306 Universit\'e Lyon 1-CNRS, Universit\'e de Lyon 69622 Villeurbanne, France}

\cortext[cor1]{ Corresponding author}

\begin{abstract}
Recent experimental and theoretical investigations
of crystal growth from solution in the vicinity of an impermeable wall
 have shown that: (i) growth can be maintained within the contact region 
when a liquid film is present between the crystal and the substrate;
(ii)  a cavity can form in the center of the contact region due to insufficient supply of mass through the liquid film.
Here, we investigate the influence of surface kinetics on these phenomena
using a thin film model. 
First, we determine the growth rate within the confined region in the absence of a cavity. 
Growth within the contact induces a drift of the crystal away from the substrate. 
Our results suggest novel strategies to measure surface kinetic coefficients 
based on the observation of this drift.
For the specific case where growth is controlled by 
surface kinetics outside the contact, we show that the total displacement of the crystal due to the growth in the contact is finite. As a consequence, the  
growth shape approaches asymptotically the free growth shape truncated 
by a plane passing through the center of the crystal. 
Second, we investigate the conditions under which a cavity forms. 
The critical supersaturation above which the cavity forms is 
found to be larger for slower surface kinetics. 
In addition, the critical supersaturation decays as a power law of the contact size. 
The asymptotic value of the critical supersaturation and the exponent of the decay are 
found to be different for attractive and repulsive disjoining pressures. 
Finally, our previous representation of the transition
within a morphology diagram appears to be uninformative in the limit of slow surface kinetics.
\end{abstract}

\begin{keyword}
A1.~Growth models\sep A1.~Thin film model \sep A2.~Growth from solution \sep A1.~Confinement \sep A1.~Morphological stability\sep A1.~Surface kinetics
\end{keyword}

\end{frontmatter}

%%%%%%%%%%%%%%%%%%%%%%%%%%%%%%%%%%%%%%%%%%%%%%%%%%%%%%%%%%%%%%%%%%%%%%%%%%
%%%%%%%%%%%%%%%%%%%%%%%%%%%%%%%%%%%%%%%%%%%%%%%%%%%%%%%%%%%%%%%%%%%%%%%%%%
%%%%%%%%%%%%%%%%%%%%%%%%%%%%%%%%%%%%%%%%%%%%%%%%%%%%%%%%%%%%%%%%%%%%%%%%%%

\section{Introduction}\label{sec:intro}

\noindent  
Crystal growth often occurs in the
vicinity of substrates.
In solution, growth on a substrate
arises for example after sedimentation~\cite{Biscaye1965,Salvarezza1996} 
or after heterogeneous nucleation~\cite{Markov2016,Winter2009,Page2006,Chayen2006}.
When the substrate is impermeable, growth can still occur at the surface of the crystal facing the substrate
if a liquid film is present between the crystal and 
the substrate.
Recent theoretical and experimental studies~\cite{Kohler2018,Gagliardi2018a}
have pointed out that in these conditions,
a cavity can form on the confined crystal surface.  
The cavity forms due to an insufficient supply of growth units in the center of the contact.
Indeed mass transport along the liquid film is limited due to the smallness of the film thickness.
After its formation, the cavity expands and gives rise to growth rims around 
the contact region which have been observed since the beginning of the 20th century~\cite{Becker1905,Taber1916}, 
and have also attracted recent interest~\cite{Royne2012,Li2017}.

The aim of this paper is to discuss the influence of surface kinetics
on the growth of crystals that are located in the vicinity of a flat substrate. 
We start in \cref{sec:crossover,sec:thin_film_model} by extending the thin film model introduced in Ref.~\cite{Gagliardi2018}
to account for slow surface kinetics. 

Using this model, we show in \cref{sec:lifting} that surface
kinetics influences the rate of growth within the contact region when the contact 
width is smaller than a critical length scale $l_0=(Dh/\nu)^{1/2}$,
where $D$ is the diffusion constant of growth units in the liquid film,
$h$ is the thickness of the film, and $\nu$ is the surface kinetics
coefficient. A quantitative prediction for the growth rate is obtained in the case of disjoining pressures
which exhibit an attractive tail at long distance. 
This situation occurs for nanometer-scale film thicknesses, 
where attractive van der Waals interactions are relevant. 
A similar trend is observed for repulsive disjoining pressures.
We also analyze the asymptotic growth shape when
the growth outside the contact is controlled by surface kinetics.
Growth within the contact induces a drift of the crystal bulk
away from the substrate. The total displacement resulting from this 
drift is found to be finite. As a consequence, the asymptotic
growth shape is the free growth shape truncated by
a plane passing through the center of the crystal.

In \cref{sec:cavity_formation}, we discuss the conditions under which a cavity forms
within the contact in the presence of slow surface kinetics. 
For large contact sizes, the 
critical supersaturation above which a cavity forms
vanishes in the case of repulsive interactions, while it
reaches a constant value for attractive interactions.
This statement extends the results of Ref.~\cite{Gagliardi2018a}
to the case of slow kinetics.
The power-law behavior characterizing the decay of the critical supersaturation
toward its asymptotic value depends on surface kinetics and on the type of interaction.
Finally, a straightforward generalization of the morphology diagram proposed in Ref.~\cite{Kohler2018,Gagliardi2018a}
is found to be uninformative for slow kinetics, despite good data collapse.

%%%%%%%%%%%%%%%%%%%%%%%%%%%%%%%%%%%%%%%%%%%%%%%%%%%%%%%%%%%%%%%%%%%%%%%%%%
%%%%%%%%%%%%%%%%%%%%%%%%%%%%%%%%%%%%%%%%%%%%%%%%%%%%%%%%%%%%%%%%%%%%%%%%%%
%%%%%%%%%%%%%%%%%%%%%%%%%%%%%%%%%%%%%%%%%%%%%%%%%%%%%%%%%%%%%%%%%%%%%%%%%%

\section{Surface kinetics}
\label{sec:crossover}

\noindent Consider a rigid crystal, where elastic deformations are neglected.
The local change of volume of the crystal at the interface is the difference
between the normal velocity of the interface $v_n$ and
the projection of the rigid-body crystal velocity $\mathbf u$ on the surface normal $\hat{\mathbf{n}}$. 
Since the crystal is rigid, the crystallization rate $\mathrm v$ is simply proportional 
to this local change of volume at the surface:
\begin{align}
\label{eq:growth_rate_def}
\mathrm v=v_n-\hat{\mathbf{n}}\cdot\mathbf{u}\, .
\end{align}
We assume a linear kinetic law, which relates the crystallization rate to the
departure from equilibrium.
This is measured by the difference between the concentration $c$ 
of crystal molecules in the liquid in front of the crystal
and its equilibrium value $c_{eq}$:
\begin{equation}
\label{eq:kin_law_general_main}
\mathrm v= \Omega\nu (c-c_{eq})\, ,
\end{equation}
where $\Omega$ is the molecular volume of the solid crystal, and $\nu$ a kinetic constant.

\begin{figure}
\includegraphics[width=\linewidth]{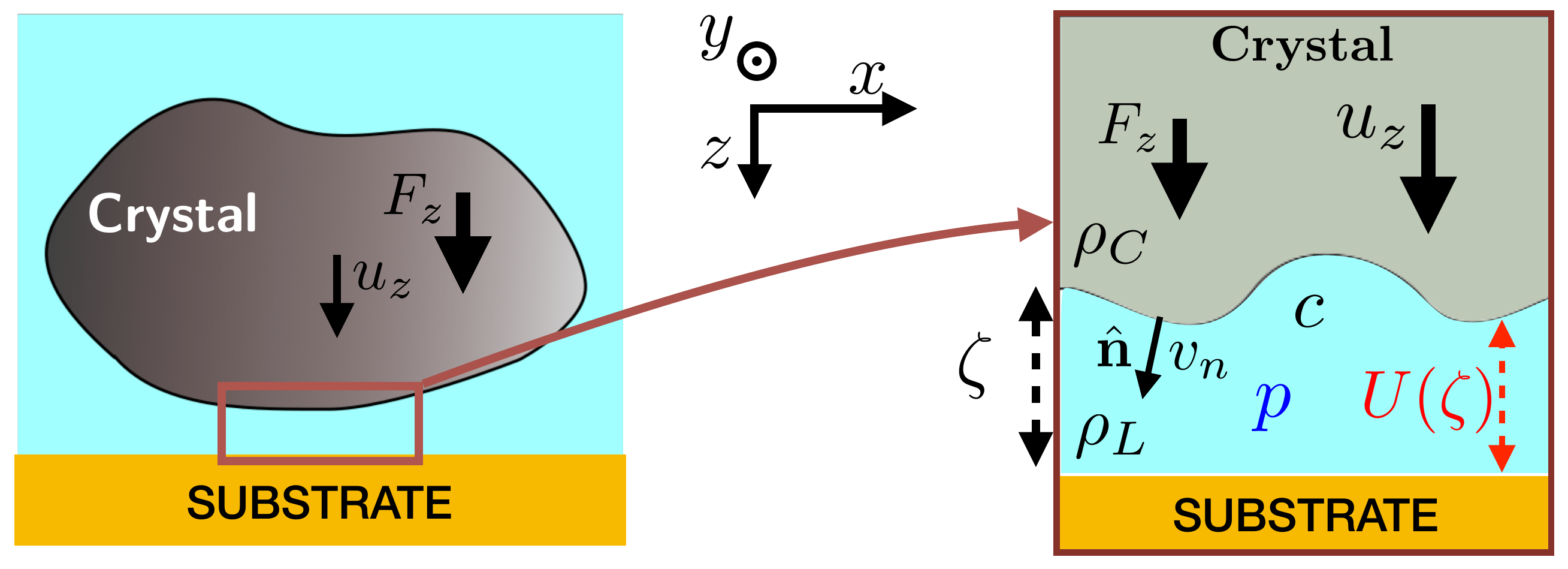}
\caption{Sketch of a crystal in the vicinity of a substrate, and zoom in the contact region. 
Notations are defined in the text and in \cref{app:model}. 
\label{fig:sketch}}
\end{figure}

Let us now assume that the crystal is growing in the vicinity of an impermeable and
flat substrate.  
A schematics of the system is presented in \cref{fig:sketch}. 
Neglecting hydrodynamic advection, we assume that mass transport of growth units 
within the liquid film is controlled by diffusion. From Fick's law, the local
diffusion flux in the liquid is $-D\nabla c$. When the film is thin,
diffusion in the $z$-direction orthogonal to the substrate
leads to fast relaxation of the concentration to a value that does not
depend on $z$~\cite{Gagliardi2018}. As a consequence,
the total mass flux in the directions $x,y$ parallel to the substrate
is simply $-D\zeta\nabla_{xy} c$, where $\zeta(x,y)$ is the local thickness of the thin liquid film between
the crystal and the substrate, and  $\nabla_{xy}=(\partial_x,\partial_y)$ is the gradient
operator in the $x,y$ plane. From mass conservation, the  divergence of the diffusion flux must be proportional to the crystallization rate
\begin{align}
\label{eq:mass_cons_lubric}
\mathrm{v}\approx \Omega\nabla_{xy}\cdot[ D\zeta\nabla_{xy} c],
\end{align}
where $\Omega$ is the molecular volume of the crystal.
(This latter equation can be derived formally from the lubrication
expansion in the dilute limit, as discussed in \cref{app:model}).
Assuming that the variations of the concentration and thickness parallel
to the substrate occur at a length scale $\ell$, \cref{eq:mass_cons_lubric} indicates that
the crystallization rate must be of the order of $\Omega D\zeta c/\ell^2$. 
Comparing this expression with that enforced by
surface kinetics \cref{eq:kin_law_general_main}, we obtain a lengthscale
\begin{align}
\label{eq:diff_length}
l_0\approx \sqrt{\frac{Dh}{\nu}} \, ,
\end{align}
where we assumed that $\zeta\approx h$, 
the typical width of the film.  %(generally enforced by the interaction with the substrate, see below).

The dynamics is in general controlled by the slowest process. As a consequence,
we will obtain a surface-kinetics-limited regime
at scales $\ell\ll l_0$, and a diffusion-limited regime
at scales $\ell\gg l_0$.

Unfortunately, precise and reliable experimental measurements
of kinetic constants are scarce, and quantitative values of $\nu$ 
reported in the literature can be very different for the same material~\cite{Colombani2012}. 
For instance, the reported kinetic constants $\nu$ range from $10^{-7}$ to $10^{-4}$ms$^{-1}$~\cite{Colombani2016,Plummer1976} for calcite, 
and from $\nu\sim10^{-5}$ to $10^{-3}$ms$^{-1}$~\cite{Naillon2017} for salt.

Let us consider the specific case of the 
experiments of Kohler {\it et al} \cite{Kohler2018} with NaClO$_3$.
We obtain an upper bound for $l_0$
when using the smallest reported values for the kinetic constant $\nu\sim 10^{-5}$ ms$^{-1}$~\cite{Wang1996}, 
a large diffusion constant $D\sim 10^{-9}$m$^2$s$^{-1}$ 
(at saturation we rather expect $D\sim10^{-10}\mathrm{m^2 s^{-1}}$~\cite{Campbell1969,Kohler2018} )
%[{\color{blue}We have to be careful: for NaCl at saturation $10^{-9}\mathrm{m^2 s^{-1}}$ has been reported~\cite{Chang1985}}], 
and a large thickness $h\sim 100$nm, which is upper bound for the experimental setup
of Kohler {\it et al} \cite{Kohler2018}. This leads to $l_0\sim 3\mu$m.
Such a value of $l_0$ is small as compared to the crystal sizes used in~Ref.\cite{Kohler2018} 
which were ranging from $10\mathrm{\mu m}$ to $100\mathrm{\mu m}$. 
Hence, we can safely assume that in those experiments 
the formation of the cavity was indeed within the diffusion-limited regime.

However, if we consider a less soluble material such as CaCO$_3$, 
characterized by a much slower surface kinetics, the scenario could be different. 
Using again $D\sim 10^{-9}\mathrm{m^2s^{-1}}$ and $h\sim100$nm and assuming 
the smallest kinetic constant reported $\nu=10^{-7}$ms$^{-1}$~\cite{Colombani2016}, 
we obtain $l_0\sim 30\mathrm{\mu m}$. 
This value is comparable to the crystal sizes used in experiments~\cite{Kohler2018,Li2017}.

%%%%%%%%%%%%%%%%%%%%%%%%%%%%%%%%%%%%%%%%%%%%%%%%%%%%%%%%%%%%%%%%%%%%%%%%%%
%%%%%%%%%%%%%%%%%%%%%%%%%%%%%%%%%%%%%%%%%%%%%%%%%%%%%%%%%%%%%%%%%%%%%%%%%%
%%%%%%%%%%%%%%%%%%%%%%%%%%%%%%%%%%%%%%%%%%%%%%%%%%%%%%%%%%%%%%%%%%%%%%%%%%
\section{Thin film Model}
\label{sec:thin_film_model}
%%%%%%%%%%%%%%%%%%%%%%%%%%%%%%%%%%%%%%%%%%%%%%%%%%%%%%%%%%%%%%%%%%%%%%%%%%
\subsection{Model equations}
\label{sec:model_equations}

\noindent Following the same lines as in Ref.~\cite{Gagliardi2018},
the dynamics within the contact is described 
by a thin film model based on the small slope limit (also called the lubrication limit)~\cite{Oron1997}.
Details about the derivation of these 
equations are reported in \cref{app:model}. In the following,
we only provide a heuristic discussion of the 
resulting equations.

The model assumes a rigid crystal, equal solid and liquid densities and the dilute limit.
Rotations and translations along the plane of the substrate $x,y$ are not considered.
The motion of the solid is restricted to translations
along the $z$ axis. Hence, the rigid body crystal velocity is $\mathbf u=u_z\hat{\mathbf z}$, where $\hat{\mathbf z}$ is the unit vector along the $z$ axis.

In the lubrication limit, the slopes are small, i.e., $|\nabla \zeta |\ll 1$. As a
consequence,  
the local growth rate is approximated by the rate along the $z$ direction:
$\mathrm{v}\approx \mathrm{v}_z$.
We focus on the case of an axisymmetric contact,
while general equations  in non-axisymmetric geometries are reported in \cref{app:model}.
In the axisymmetric geometry, \cref{eq:growth_rate_def,eq:kin_law_general_main,eq:mass_cons_lubric} read
\begin{align}
\label{eq:def_vz}
\partial_t\zeta(r,t) &= -\mathrm{v}_z(r,t) - u_z(t)\, ,
\\
\label{eq:concentration}
c(r,t) &= c_{eq}(r,t) + \frac{\mathrm{v}_z(r,t)}{\Omega \nu}\, ,
\\
\label{eq:main}
\mathrm{v}_z(r,t) &= D\Omega\frac{1}{r}\partial_r\Bigl[r\zeta \partial_r c(r,t) \Bigr] \, ,
\end{align}
with $D$ the diffusion constant, $\Omega$ the molecular volume 
in the crystal and $c$ the concentration at the liquid-crystal interface.

In the dilute limit, the local equilibrium concentration $c_{eq}$ entering in \cref{eq:concentration}
depends on the local chemical potential $\Delta\mu$
via the thermodynamic relation
\begin{align}
\label{eq:ceq}
c_{eq}  &= c_0\exp\left[\frac{\Delta\mu}{k_BT}\right]\approx c_0\left(1+\frac{\Delta\mu}{k_BT}\right)
\end{align}
where $c_0$ is the solubility, $k_B$ the Boltzmann constant, 
$T$ the temperature (assumed constant and homogeneous), 
and $\Delta\mu$ is the chemical potential at the crystal-liquid interface. 
Following \cite{Gagliardi2018,Kohler2018}, we assume $\Delta\mu/k_BT\ll 1$ 
and we linearize the exponential in \cref{eq:ceq}. 
Accounting 
for anisotropy and for the presence of a substrate~\cite{Gagliardi2018,Olivier2016,deGennes2004},
the chemical potential is composed of two contributions
\begin{align}
\label{eq:mu_main}
\frac{\Delta\mu}{\Omega} &= \tilde{\gamma}\kappa - U'(\zeta)
\end{align}
The first term in \cref{eq:mu_main}  depends on the surface free-energy $\gamma$.
Since we assume an axisymmetric crystal, the surface free energy $\gamma(\theta)$
depends only on the angle $\theta=\arctan\partial_r\zeta$.
This term is proportional to the surface stiffness~\cite{Saito1996} 
$\tilde{\gamma} = \gamma(0) + \gamma''(0)$. In addition, the local mean curvature in cylindrical coordinates reads
\begin{align}
\kappa = \partial_{rr}\zeta + \partial_r\zeta/r\, .
\end{align}
The second term in the right-hand-side of \cref{eq:mu_main} accounts for disjoining pressure effects, 
where $U(\zeta)$  is the interaction potential between the crystal and the substrate
(i.e., the free energy cost per unit area for reducing the 
film thickness from a large value to its actual value $\zeta$).

Combining \cref{eq:concentration,eq:main,eq:ceq} we obtain:
\begin{equation}
\label{eq:evolution_vz}
\mathrm{v}_{z}=\frac{1}{r}\partial_r\Bigl[r\zeta B\partial_r (\tilde{\gamma}\kappa - U'(\zeta)) \Bigr] 
+\frac{1}{r}\partial_r\left[r\frac{\zeta D}{\nu}\partial_r \mathrm{v}_{z}\right]\, .
\end{equation}
where $B={\Omega^2 Dc_0}/({k_BT})$.

Finally, the force balance between an external force $F_z$, 
viscous dissipation, and disjoining pressure provides an additional relations
which allows one to determine $u_z$~\cite{Gagliardi2018}
\begin{equation}
\label{eq:uz}
u_{z} \, 2\pi\int_0^R \!\!\!\! \mathrm{d}r\, r\int_r^{R}\!\!\!\! \mathrm{d}r'\, \frac{6\eta r'}{\zeta(r')^3} 
= 
F_{z} + 2\pi\int_0^R\!\!\!\! \mathrm{d}r \, r U'(\zeta)\, ,
\end{equation}
where $\eta$ is the liquid viscosity.

The system \cref{eq:def_vz,eq:evolution_vz,eq:uz} 
provide a closed set of equations for the evolution of $\zeta(r,t)$, and $u_{z}(t)$.

%%%%%%%%%%%%%%%%%%%%%%%%%%%%%%%%%%%%%%%%%%%%%%%%%%%%%%%%%%%%%%%%%%%%%%%%%%
\subsection{Disjoining pressure and sedimentation force}

\noindent The disjoining pressure $U'$ is due to the interactions between surfaces immersed in liquid. 
This pressure is usually modeled by the DLVO theory~\cite{Verwey1948,Israelachvili2011}.
However, additional short-range forces are often present, leading to a wide variety
of possible dependences of the disjoining pressure on the thickness.
Here, we consider two prototypical cases, both for the sake of clarity
and to allow for direct comparison with our previous works~\cite{Kohler2018,Gagliardi2018a}.

The first potential is purely repulsive and is aimed to mimic
the effect of protrusions of the substrate surface 
or particles located between the crystal and the substrate,
that cannot be engulfed in the crystal. Due to these impurities,
the crystals cannot approach the substrate at distances smaller than 
a minimal thickness $h$. 
In the experiments of Ref.~\cite{Kohler2018},
$h$ varied from $10\mathrm{nm}$ to $100$nm. 
In order to account for this minimal thickness,
we choose a potential $U(\zeta)$ that diverges
when $\zeta\rightarrow h$
\begin{equation}
\label{eq:potential_Felix}
U(\zeta) = \bar{a} A \frac{e^{\frac{-(\zeta-h)}{h\bar{\lambda}}}}{\zeta-h}\, ,
\end{equation}
where $h\bar{\lambda}$ is a decay length and $\bar{a}$ a dimensionless interaction amplitude.

This repulsive potential is accompanied by a buoyancy sedimentation force 
$F_z = F_g  = \Delta\rho g(2R)^3$,
where $g$ is the gravitational acceleration, and $\Delta \rho$
is the solid-liquid density difference. 
Such a sedimentation force maintains the crystal in the vicinity of the substrate\footnote{
Note that, even though contributions due to density differences are neglected in the chemical potential~\cite{Gagliardi2018}, 
they are kept to compute the external gravitational force maintaining the crystal 
close to the substrate in the case of a purely repulsive interaction.}.
 In Ref.~\cite{Kohler2018}, the results of the model were shown to be insensitive 
to the numerical value of $F_g$ and $\bar{a}$. 
Thus, irrespective of the nature of the material, 
we compute the gravitational force 
with the parameters corresponding NaClO$_3$ 
and use $\bar{a} = 10^{-3}$. 
We also use a small interaction range $\bar{\lambda}=10^{-2}$
to ensure a film thickness close to $h$.

The second type of interaction potential accounts for smaller
distances between the crystal and the substrate $\zeta\leq 10 $nm.
At these distances, van der Waals attractive
forces cannot be neglected~\cite{Israelachvili2011}.
We therefore combine  
a long-range van der Waals attraction $\sim \zeta^{-2}$
with a a shorter range repulsion $\sim \zeta^{-3}$:
\begin{equation}
\label{eq:potential_sub}
U(\zeta) = \frac{A}{12\pi}\Bigl( -\frac{1}{\zeta^2} +\frac{2h}{3\zeta^3} \Bigr)\, .
\end{equation}
Note that in this case $h$ corresponds to the position 
of the minimum of the potential well.

Since the energy cost for the formation of a surface at $\zeta=h$ is lower than 
the energy cost far away from the substrate at $\zeta\rightarrow\infty$,
heterogeneous nucleation in the vicinity of the substrate is favored with the above potential.
Hence, our study of growth with the attractive potential, \cref{eq:potential_sub}, could describe the growth of a crystal
on a substrate after heterogeneous nucleation.
Furthermore, we expect gravitational effects to be small as compared to the van der Waals attraction
at these scales. As a consequence, we will neglect the sedimentation force $F_g = 0$
when considering the potential \cref{eq:potential_sub}.

\begin{figure*}
\center
\caption{Time evolution of an axisymmetric crystal profile projected along $r$ (shaded gray) growing from solution on a substrate (yellow area) as obtained from a simulation. The interaction with the substrate is attractive, \cref{eq:potential_sub}. 
The normalized kinetic constant is $\bar{\nu} = 0.1$. 
The normalized supersaturation outside the contact area, the simulation box radius and the viscosity are $\bar{\sigma}_{bc} = 0.315$, $\bar{R} = 35$, and $\bar{\eta} = 10^{-6}$,  respectively.
For instance, assuming a typical film thickness $h=1$nm, temperature $T\approx300$K and interaction amplitude $A/6\pi\approx10^{-21}$J~\cite{Israelachvili2011,Gagliardi2018a}, and using parameters for salt~\cite{Chang1985,Naillon2017} -- $c_0 \approx 10^{27}$, $D\approx10^{-9}\mathrm{m^2s^{-1}}$, $\Omega = 45$\AA$^3$, and $\tilde{\gamma} \sim \gamma \approx 100$mJ -- we have in physical units: $R \approx 350\mathrm{n m}$, $\sigma_{bc} \approx 3.5\times 10^{-3}$ , $t \approx 10^{-4}\mathrm{s}\times \bar{t}$, $\eta \approx 10^{-3} $mPas, and $\nu = 10^{-3}\mathrm{ms^{-1}}$.
%[{\color{blue} What if $h = 10$nm ?: 
%$l = 1\mu m$, $R = 35\mu m$, $\nu = 10^{-5}m/s$,$\sigma = 3.5\times 10^{-7}$, $t = 2\times 10^3$s}
%].
 %Calcite~\cite{Gagliardi2018a,Colombani2016}, $c_0 = 10^{25}$, $D\approx10^{-9}\mathrm{m^2s^{-1}}$, $\Omega = 59$\AA$^3$, $\tilde{\gamma} \sim \gamma \approx 100$mJ, this would correspond in physical units to: $R = 350\mathrm{n m}$, $\sigma \approx 4.4\times 10^{-2}$ , $t \approx 10^{-2}\mathrm{s}\times \bar{t}$, $\eta \approx 0.1 $mPas, and $\nu = 10^{-3}\mathrm{ms^{-1}}$. 
%$R=35\mathrm{\mu m}$, $\sigma \approx 4.5\times 10^{-6}$ , $t \approx 10^{5}\mathrm{s}\times \bar{t}$, $\eta =10 $mPas, and $\nu = 10^{-6}\mathrm{ms^{-1}}$. 
%Low values for the supersaturations can be linked to the low anisotropy of the model. This issue, extensively discussed in~\cite{Kohler2018,Gagliardi2018a}, can be solved by assuming effective values for the stiffness constant $\tilde{\gamma}$. 
\label{fig:cavity} 
}
\includegraphics[width=\linewidth]{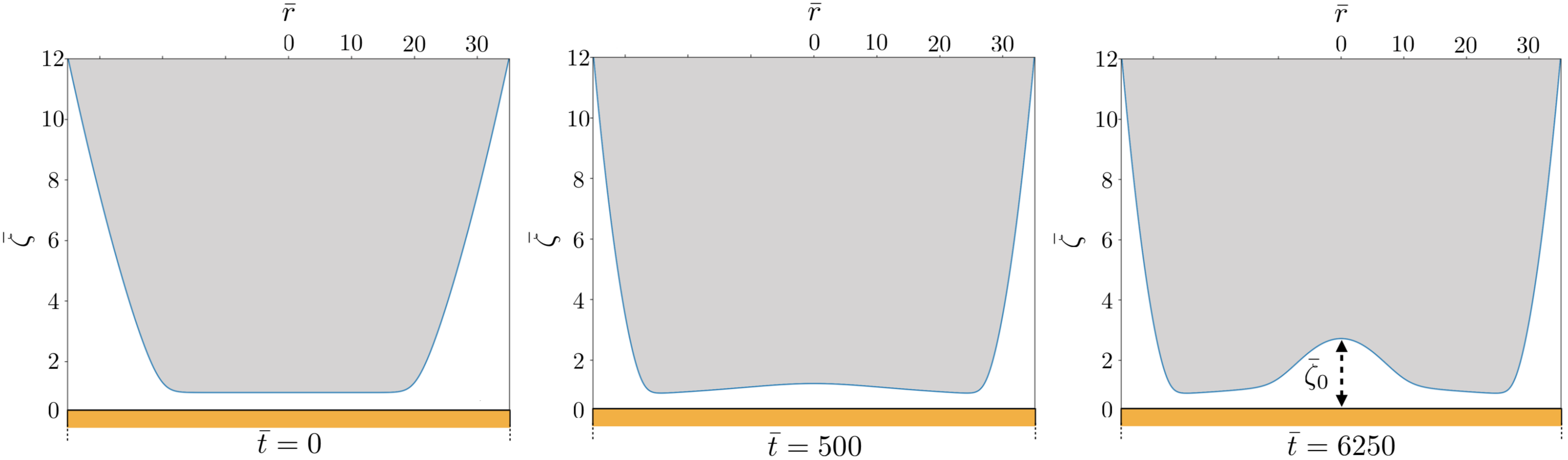}
\end{figure*}

%%%%%%%%%%%%%%%%%%%%%%%%%%%%%%%%%%%%%%%%%%%%%%%%%%%%%%%%%%%%%%%%%%%%%%%%%%
\subsection{Boundary conditions}
 \noindent We consider an integration domain (simulation box) of fixed radius $R$. 
Outside the integration domain, we assume a constant concentration and a constant pressure.
We thus assume a constant supersaturation
at the boundary of the simulation box
\begin{align}
\sigma(R) = \sigma_{bc}\, ,
\end{align}
where $\sigma(r) = c(r)/c_0-1$ with $c_0$ the solubility.
In addition we fix the film width at the boundary of the integration domain:
\begin{align}
\zeta(R) = \zeta_{bc}\, .
\end{align}
As discussed in our previous studies~\cite{Kohler2018,Gagliardi2018,Gagliardi2018a},
the model results are insensitive to the choice of $\zeta_{bc}$ when this
quantity is large enough to ensure vanishing disjoining forces at the
boundary of the integration domain $U'(\zeta_{bc})\approx 0$.

%%%%%%%%%%%%%%%%%%%%%%%%%%%%%%%%%%%%%%%%%%%%%%%%%%%%%%%%%%%%%%%%%%%%%%%%%%
\subsection{Normalization of model equations}
\label{sec:normalization}

\noindent \Cref{eq:evolution_vz,eq:uz} are solved in normalized units.
We start defining a dimensionless repulsion strength  
$\bar{A} = A/(\tilde{\gamma} h)$ for the repulsive interaction \cref{eq:potential_Felix}, 
and $\bar{A} = A/(6\pi\tilde{\gamma} h^2)$ for the attractive one \cref{eq:potential_sub}.
%$\bar{A} = A/(\tilde{\gamma} h)$, the normalized film thickness 
Moreover, the normalized film thickness and radial coordinate are defined as $\bar{\zeta} = \zeta / h$ and $\bar{r} = r/l $, 
where $l = h/\bar{A}^{1/2}$. 
The normalized time variable is  
$\bar{t} = B \tilde{\gamma} h t/l^4$. 
Other relevant dimensionless quantities are the normalized system size
\begin{equation}
\label{eq:norm_size}
 \bar{R}=\frac{R}{l}\, , 
\end{equation}
the normalized supersaturation 
\begin{equation}
\label{eq:scaling_sup}
\bar{\sigma} = \frac{k_BT l^2}{\Omega\tilde{\gamma} h} \sigma \, ,
\end{equation}
the normalized force
\begin{align}
\bar{F}_{z} = \frac{F_{z}}{\tilde{\gamma} h},
\end{align}
and the dimensionless vertical rigid body velocity of the crystal 
\begin{equation}
\label{eq:scaling_v}
\bar{u}_{z} = \frac{l^4}{h^2 \tilde{\gamma}B} u_{z}\, .
%\frac{l^4k_BT}{h^2 \tilde{\gamma}D\Omega^2c_0} u_{z}\, ,
\end{equation}
We also define the normalized viscosity
\begin{equation}
\label{eq:scaling_eta}
\bar{\eta} = \frac{B}{h^2}\eta\, .
\end{equation}
Finally, a central dimensionless quantity is the normalized kinetic constant
\begin{equation}
\label{eq:scaling_nu}
\bar{\nu} = \frac{l^2}{hD}\nu\, .
\end{equation} 
Since the aim of this paper is to focus on 
surface kinetics, we will vary $\bar{\nu}$
while the viscosity is kept small, $\bar{\eta} = 10^{-6}$.   
Larger viscosities are known to cause additional phenomena such as the hindering of cavity formation~\cite{Gagliardi2018a},
but the cross effects of viscosity and surface kinetics are beyond the scope of this paper.

%%%%%%%%%%%%%%%%%%%%%%%%%%%%%%%%%%%%%%%%%%%%%%%%%%%%%%%%%%%%%%%%%%%%%%%%%%

%%%%%%%%%%%%%%%%%%%%%%%%%%%%%

\subsection{Numerical methods}
\label{sec:numerical_methods}

\noindent The crystal evolution was computed numerically using the following steps.
All spatial derivatives are calculated using a finite difference scheme,
with spatial discretization $\Delta\bar{r}=0.2$. 
At a given time-step $t$, the local growth rate $\mathrm{v}_z(r,t)$ is calculated from \cref{eq:evolution_vz} by matrix inversion.
In addition, force balance equation \cref{eq:uz} determines the crystal velocity $u_z(t)$. 
Then, the profile is computed at the next time step $t+\Delta t$ using forward Euler integration based on \cref{eq:def_vz}.
For the repulsive potential, \cref{eq:potential_Felix}, 
the numerical scheme proves to be stable using a time-step $\Delta\bar{t}$ from $10^{-4}$ 
for large kinetic constants ($\bar{\nu}=100$), to $ 10^{-2}$ for small kinetic constants ($\bar{\nu}<0.005$).
In the case of the attractive potential, \cref{eq:potential_sub}, the time step is kept fixed at $\Delta\bar{t} = 10^{-4}$.
Finally, the simulations presented in this work use a normalized thickness 
at the boundary of the integration domain $\bar{\zeta}(\bar{R}) = \bar{\zeta}_{bc} = 12$.

For illustrative purposes, in \cref{fig:cavity} we show the temporal evolution of a crystal profile along $r$ with a cavity gradually forming.
Eventually, all simulations reach steady-states.
We assume that the variations of the contact size are slow 
as compared to the relaxation of the profile towards steady states within the contact.
Based on this hypothesis, the profile of a growing crystal with a time-dependent contact
size $R(t)$, is approximated by a family of steady-state profiles at each $R(t)$. 
Such a \emph{quasi-static} hypothesis is a central assumption of this paper. 
This assumption has been discussed in our previous studies of the thin film model~\cite{Gagliardi2018,Gagliardi2018a,Kohler2018},
and compared favorably with experiments 
where crystals were also growing laterally~\cite{Kohler2018}.
Since we define a steady-state as a state where the film
profile does not change in time, we have $\partial_t\zeta=0$ in steady-state.
As a consequence, \cref{eq:def_vz} implies $\mathrm{v}_z=u_{z}$,
i.e. the crystallization rate $\mathrm{v}_z$ and the 
crystal velocity $u_{z}$ coincide in steady-state.

In the following, the radius $L$ of the contact region, which is smaller
than the radius $R$ of the simulation box, will be used to analyze the results
of the simulations. 
The contact radius $L$ is calculated from the heuristic definition 
$L = \max_r[\partial_r\kappa_{1D}]$, where $\kappa_{1D}=\partial_{rr}\zeta$ 
is the 1D curvature of the crystal profile. 
As discussed in \cref{app:contact_radius}, this definition provides a reasonable evaluation of the contact radius
in agreement with the different heuristic estimates used in
our previous studies Refs.~\cite{Gagliardi2018a,Kohler2018}.
We have chosen this new definition because it does not refer explicitly
to the reference film thickness $h$ and because it does not
require any ad hoc parameter. 

The analysis of the simulation results below will also require the evaluation of the
supersaturation  $\sigma_b$  
at the edge of the contact. This supersaturation
is simply defined by the relation $\sigma_b=\sigma(L)$ using \cref{eq:concentration}.

% {\color{blue}In the following, simulation results are given in normalized units. Indeed, the purpose of this work is to discuss qualitative behaviors. In any case, relevant results will be given in physical units in the text.}

%%%%%%%%%%%%%%%%%%%%%%%%%%%%%%%%%%%%%%%%%%%%%%%%%%%%%%%%%%%%%%%%%%%%%%%%%%
%%%%%%%%%%%%%%%%%%%%%%%%%%%%%%%%%%%%%%%%%%%%%%%%%%%%%%%%%%%%%%%%%%%%%%%%%%
%%%%%%%%%%%%%%%%%%%%%%%%%%%%%%%%%%%%%%%%%%%%%%%%%%%%%%%%%%%%%%%%%%%%%%%%%%

\begin{figure*}[!t]
\center
\caption{Top panels: normalized growth rate $|\bar u_z|$ in the absence of cavity 
as a function of the normalized kinetic constant $\bar\nu$.
Simulation results are indicated by filled black dots.
The red empty dots and dashed lines report 
the prediction of \cref{eq:velocity_slowKin} using
the chemical potential $\Delta\mu_b$ extracted from simulations.  %Terms of order $1/L^2$ in the denominator are neglected.
Simulation box: $\bar{R} = 40$; normalized viscosity: $\bar{\eta} = 10^{-6}$.
\textbf{a)} Attractive interaction potential, \cref{eq:potential_sub}. 
The same supersaturation $\bar{\sigma}_{bc} = 0.2$ at the edge of the simulation box is used for all simulations.
The contact area increases with $\bar{\nu}$: from $\bar{L} = 24.6$ to $\bar{L} =30$. 
The supersaturation at $\bar{L}$ decreases from $\bar{\sigma}_b\approx 0.19$ to $0.15$. 
The equilibrium chemical potential $\Delta\mu_{eq}$ is determined 
by the relation \cref{eq:chemical_pot_eq} (\cref{app:perturantion_analysis}).
\textbf{b)} Repulsive interaction, \cref{eq:potential_Felix}. Here $\Delta\mu_{eq} = 0$ 
and $\sigma_{bc}$ is chosen close to the transition for each dot in order to obtain a roughly flat
film width $\approx h$. 
First dot on the left $\bar{L} = 14.6$ and $\bar{\sigma}_b \approx 0.48$, last $\bar{L} =16.8$ and $\bar{\sigma}_b \approx 0.02$.\\
Bottom panels: Steady state crystal profiles (black solid line) as obtained by simulation. Fixed simulation parameters: $\bar{R} = 40$, $\bar{\eta} = 10^{-6}$, and $\bar{\nu} = 5\times 10^{-3}$. 
\textbf{c)}  Attractive potential, \cref{eq:potential_sub}. Red dashed line: analytic expression from linear perturbation analysis, \cref{eq:pert}. The contact size -- measured from the criterion $\bar{L}=\max_{\bar{r}} [\partial_{\bar{r}}(\partial_{\bar{r}\bar{r}}\bar{\zeta})]$ -- is $\bar{L}=30.6$ (boundary of the analytical curve).
Supersaturation at the boundary of the simulation box: $\bar{\sigma}_{bc} = 0.45$. A cavity would be observed at a critical supersaturation $\bar{\sigma}_{bc}^{cav} =0.53$.
\textbf{d)} Repulsive potential, \cref{eq:potential_Felix}.
The supersaturation, $\bar{\sigma}_{bc} \approx 0.17$, is the critical one. The contact size is $\bar{L}=16$.
%Simulation parameters are: $\bar{R} = 40$, $\bar{\sigma}_{bc} = \approx 0.17$, and $\bar{\nu} = 5\times 10^{-3}$. %The formation of a cavity would be observed at $\bar{\sigma}_{bc} =0. 53$.
\label{fig:vel_nu}}
%\begin{subfigure}[t]{0.47\linewidth}
%\includegraphics[width=\linewidth]{figures/vel_nu_attr}
%\end{subfigure}
%\hspace{2mm}
%\begin{subfigure}[t]{0.48\linewidth}
%\includegraphics[width=\linewidth]{figures/vel_nu_rep}
%\end{subfigure}
%
%\begin{subfigure}[b]{0.47\linewidth}
%\includegraphics[width=\linewidth]{figures/profile_pert}
%\end{subfigure}
%\hspace{2mm}
%\begin{subfigure}[b]{0.47\linewidth}
%\includegraphics[width=\linewidth]{figures/profile_rep}
%\end{subfigure}
\includegraphics[width=\linewidth]{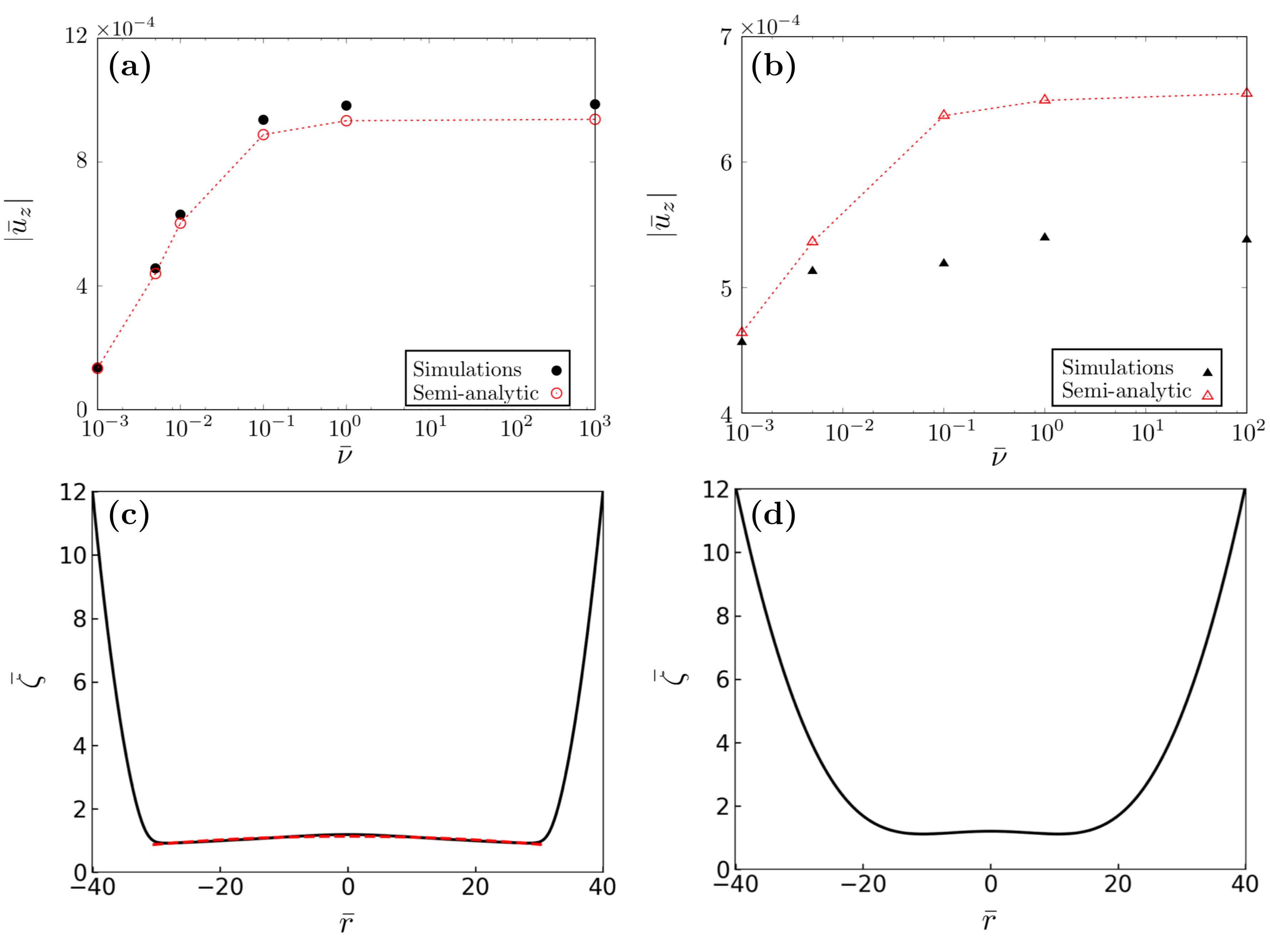}
\end{figure*}

\section{Growth before cavity formation}
\label{sec:lifting}

%%%%%%%%%%%%%%%%%%%%%%%%%%%%%%%%%%%%%%%%%%%%%%%%%%%%%%%%%%%%%%%%%%%%%%%%%%
\subsection{Growth rate}

\noindent We start by studying the rate of crystal growth in steady-state and
in the absence of a cavity.  

In the case of an attractive interaction potential~\cref{eq:potential_sub}
and in the diffusion-limited regime, a perturbative analysis around a flat equilibrium profile
($\zeta=h$) was performed in Ref.~\cite{Gagliardi2018a}.
This analysis allowed us to determine the steady-state profile of the film and crystal rigid body velocity $u_z$ (growth rate). 
We have performed a similar analysis in the case where surface kinetics
cannot be neglected. The details of the calculation, reported in  \cref{app:perturantion_analysis}, predict a concave film profile which is in reasonable agreement with simulations (see \cref{fig:vel_nu}c).
This analysis also provides an expression for the steady-state growth rate: 
\begin{equation}
\label{eq:velocity_slowKin}
|u_{z}|\approx\frac{\Omega c_0\sigma_b - \Delta\mu_{eq}/(k_BThD)}{(6\bar{\eta} + 1/2) L^2/(4hD) + \nu^{-1}}\, ,
\end{equation}
where $\Delta\mu_{eq}$ is the equilibrium chemical potential. Since $\Delta\mu_{eq}\sim 1/L$ (see \cref{app:perturantion_analysis_vel}), 
this term can be neglected for large crystals.

As illustrated in \cref{fig:vel_nu}a, simulations with an attractive interaction are in good agreement with \cref{eq:velocity_slowKin}. 
In \cref{fig:vel_nu}b we also show that for a purely repulsive interaction, \cref{eq:velocity_slowKin} can still grasp the qualitative variation of the growth rate as a function of the kinetic coefficient.
However, this expression performs poorly quantitatively, especially in the limit of large
kinetic coefficients where the growth rate is about 20\% lower than the predicted value.
One  difficulty in the comparison of the repulsive case with \cref{eq:velocity_slowKin} 
is the evaluation of the contact size $L$. As a consequence, we use simulations
for the repulsive case that are not far from the threshold for the 
formation of a cavity. This leads to a flatter film. 

These results suggest novel strategies for the 
experimental measurement of the surface kinetics coefficient $\nu$.
Indeed, if the growth rate $|u_z|$  can be measured
in experiments as a function of the contact size $L$ during growth,
then the length scale $l_0$ can be extracted from
the crossover of $|u_z|$ from a constant for $L<l_0$
to a $L^{-2}$ dependence for $L>l_0$.
Then, provided that the distance $h$ between the crystal and the substrate,
and the diffusion coefficient are known, one could extract
$\nu$ from the relation $\nu = Dh/l_0^2$. %$\nu=l_0^2/(Dh)$.
We hope that strategies based on this analysis
can help to narrow down the quantitative estimates
of surface kinetic coefficients, which have been identified
as an open issue in the recent literature~\cite{Colombani2012,Colombani2016,Naillon2017}.

%%%%%%%%%%%%%%%%%%%%%%%%%%%%%%%%%%%%%%%%%%%%%%%%%%%%%%%%%%%%%%%%%%%%%%%%%%
\subsection{Crystal growth shape in surface-limited kinetics}

\begin{figure}
\includegraphics[width=\linewidth]{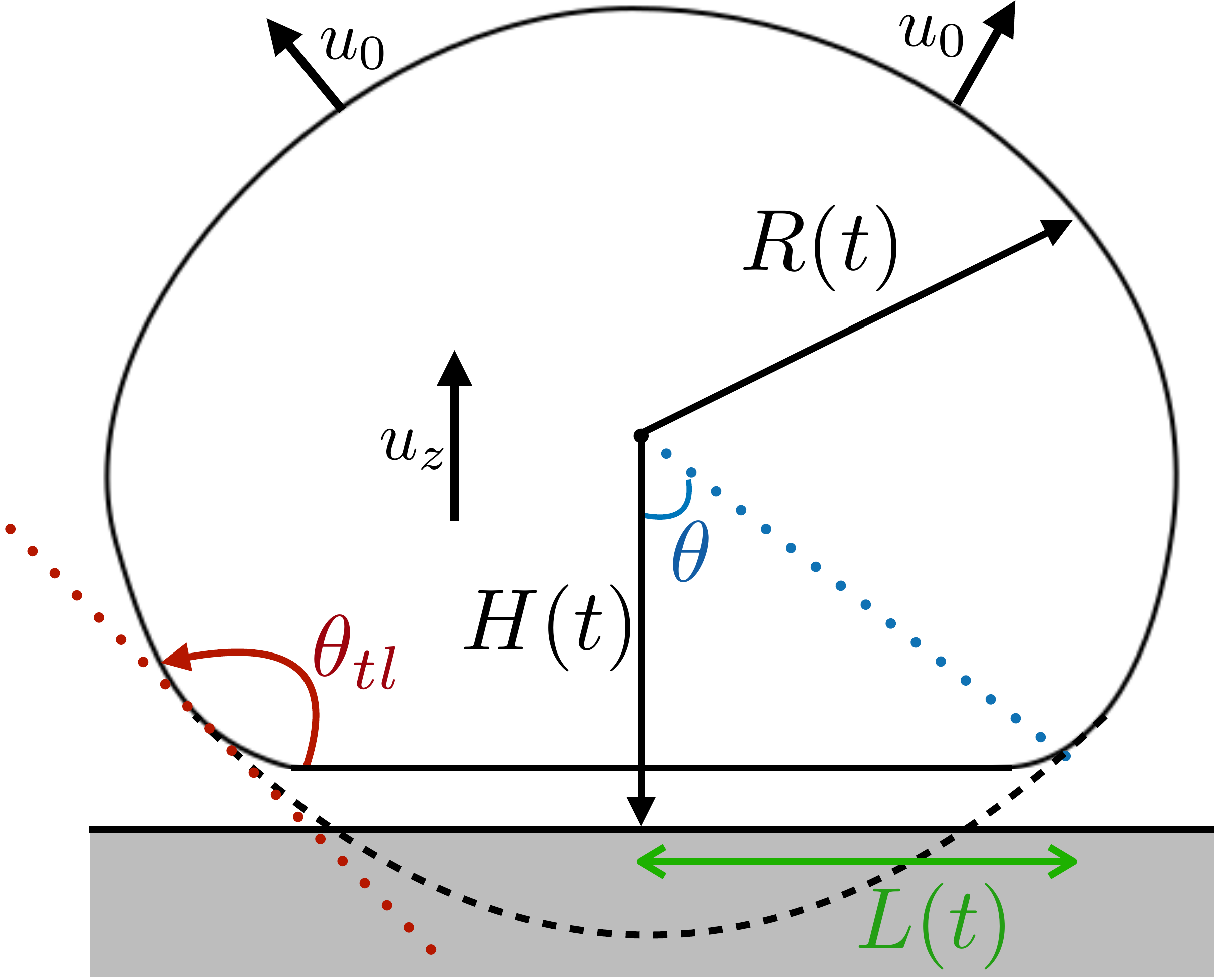}
\caption{Sketch of an isotropic crystal growing in the vicinity of a substrate, for example after heterogeneous nucleation. The evolution of the shape can be determined via \cref{eq:cryst_onSub} before a cavity appears. Notations are given in the text.
\label{fig:sketchCS}}
\end{figure}

\noindent  Here, we discuss the evolution of the global shape of the crystal, which 
results from the combination of growth within the contact and outside the contact region.
A schematic of the system is presented in \cref{fig:sketchCS}.

The expression of the quasistatic growth rate within the contact  \cref{eq:velocity_slowKin}
determines the  velocity at which the bulk of the crystal is moving away from the substrate during growth.
We focus on the case of surface-limited kinetics
outside the contact region.  
In the opposite case of diffusion-limited dynamics outside the contact, 
the determination of the growth rate outside the contact requires the assumption
of a specific far-field geometry and the full 3D solution of the diffusion equation
in this geometry. Such an analysis is beyond the scope of this paper.
In addition, we assume that surface kinetics is isotropic.  

With these assumptions, the growth rate outside the contact is expected to be constant
\[
u_0 = \nu\Omega c_0\sigma\, ,
\] 
with $\sigma$ the bulk supersaturation outside the contact area. 
The growth rate $u_0$ provides the normal velocity of the surface in the referential
moving with the bulk of the crystal. The Frank theorem~\cite{Saito1996,Pimpinelli1998} states 
that the asymptotic growth shape outside the contact is a portion of a sphere, with center $O$ and radius $R(t)=R_0+u_0t$.

Simultaneously, the shape of the crystal and the 
growth rate $u_{z}$ within the contact are assumed to relax toward the quasistatic steady-state described in the previous subsection, with the steady-state growth rate  \cref{eq:velocity_slowKin}.
The distance $H(t)$ between the center $O$ of the sphere and the substrate
therefore obeys $\dot H= |u_{z}|$.
In addition, the asymptotic contact size $L(t)$ obeys the relation $L^2 = R^2-H^2$.

Combining the above relations and using \cref{eq:velocity_slowKin} in the limit of a macroscopic crystal ($\Delta \mu_{eq} \approx 0$), 
we obtain an evolution equation for $H$
\begin{subequations}
\label{eq:cryst_onSub}
\begin{align}
\dot{H}(t)&=\frac{u_0l_0^2}{(\frac{3}{2}\bar{\eta}+\frac{1}{8})\left(R(t)^2-H^2(t)\right) + l_0^2}\, ,\\
R(t) &= R_0+u_0 t
% \\
% \theta_{tl} &=\pi -\arctan\left(\frac{R^2}{H^2}-1\right)^{1/2}
\, .
\end{align}
\end{subequations}
As discussed in \cref{app:finite_lifting}, since $R^2(t\rightarrow\infty)\gg 1$ for large integration times, 
the distance $H$ will reach a finite value $H_\infty$ as $t\rightarrow\infty$:
\begin{align}
\label{eq:h_inftty}
{H}_\infty \approx 
l_1\mathrm{Erfi}^{-1}\left[
\mathrm{Erfi}\left[\frac{H_0}{l_1}\right]
+\frac{2}{\sqrt{\pi}}
\frac{l_1{\mathrm e}^{H_0^2/l_1^2}}{R_0-H_0}
\right],
\end{align}
where
\begin{align}
l_1=\frac{l_0}{(\frac{3}{2}\bar{\eta} + \frac{1}{8})^{1/2}}.
\end{align}
The fact that $H_\infty$ is finite is one of the central results of this paper.

In the case of heterogeneous nucleation, the 
initial shape of the crystal is an equilibrium shape at the critical radius $R_c$.
As a consequence, $R_0=R_c$ and $H_0=-R_c\cos(\theta_{tl}^{eq})$ where $\theta_{tl}^{eq}$
is the equilibrium contact angle.
During the dynamics, the contact angle reads
\begin{align}
\theta_{tl} &=\pi -\arctan\left(\frac{R^2}{H^2}-1\right)^{1/2}\!\!\!\!\! .
\end{align}
Since $H_\infty$ is finite and $R$ diverges at long times,
we have $R/H\rightarrow 0$ as $t\rightarrow\infty$. It follows that 
the asymptotic contact angle does not depend on the details
of the dynamics and always converges to the same value $\theta_{tl}\rightarrow \pi/2$.

As a summary, the combination of the growth velocities outside and inside the contact 
leads to a generalization of the Frank construction for the asymptotic growth shape~\cite{Saito1996,Pimpinelli1998} which accounts for the growth in the contact with the substrate.
Since the growth rate $u_{z}$ in the contact vanishes for large crystals, 
the asymptotic contact angle is $\theta_{tl}\rightarrow \pi/2$.
These results are straightforwardly generalized for anisotropic surface-limited growth outside the contact. 
Since $H_\infty$ is finite, the asymptotic Frank shape will always be half of the free asymptotic shape, 
i.e.\ the Frank shape truncated by a plane passing through its center.
 
However, this description, valid for a flat crystal, will break down if a cavity appears in the contact. 
Below, we discuss the conditions under which a cavity will form.

%%%%%%%%%%%%%%%%%%%%%%%%%%%%%%%%%%%%%%%%%%%%%%%%%%%%%%%%%%%%%%%%%%%%%%%%%%
%%%%%%%%%%%%%%%%%%%%%%%%%%%%%%%%%%%%%%%%%%%%%%%%%%%%%%%%%%%%%%%%%%%%%%%%%%
%%%%%%%%%%%%%%%%%%%%%%%%%%%%%%%%%%%%%%%%%%%%%%%%%%%%%%%%%%%%%%%%%%%%%%%%%%
\section{Cavity formation with surface kinetics}
\label{sec:cavity_formation}
%%%%%%%%%%%%%%%%%%%%%%%%%%%%%%%%%%%%%%%%%%%%%%%%%%%%%%%%%%%%%%%%%%%%%%%%%%
%
%[{\color{red} Show also diagram with different slopes?}]
\subsection{Previous results}

\noindent To our knowledge, the first theoretical work to discuss the formation of a cavity in the contact is the seminal work of Weyl~\cite{Weyl1959}. However, this work focuses on the case where an external load is present. 
In contrast, experiments and simulations~\cite{Kohler2018,Gagliardi2018a} have recently shown that a cavity
appears in the contact region when the size of the contact or the supersaturation exceed a critical value,
in the absence of any loading force. In  the diffusion-limited case
where surface kinetics is fast enough,
the condition for the formation of a cavity was found to be~\cite{Kohler2018,Gagliardi2018a}
\begin{equation}
\label{eq:transition_fast}
|u_{z}|>|u_{z}^{cav}| = \frac{\Omega c_0 \sigma_b^{cav}}{ \alpha L_{cav}^2/(4hD)}\, ,
\end{equation}
where the index $cav$ indicates critical values at the threshold of cavity 
formation and $\alpha$ is a phenomenological constant.
Based on this relation, a morphology diagram
can be obtained by plotting the physical growth conditions
in the  plane ($|u_{z}|,\Omega 4hDc_0 \sigma_b/(\alpha L^2)$).
In this plane, the transition line is a straight line of slope one passing through the origin.
Points above this line correspond to physical situations without cavity,
and points below the line correspond to
situations with a cavity.

We wish to extend this discussion to the case where surface kinetics is slow.
To our knowledge, the only discussion of the effect of surface kinetics
on cavity formation is  Ref.~\cite{Royne2012}. This study suggests but does 
not observe directly that rims should not form for crystal sizes comparable to $l_0$.

\subsection{Heuristic derivation of the onset of cavity formation}
\label{sec:morph_D_derivation}
\begin{figure*}
\center
\caption{Two alternative representations of a non-equilibrium morphology diagrams summarizing the threshold for cavity formation at different normalized surface kinetic constants, $\bar{\nu}$. 
The phenomenological constant is $\alpha = 0.61$. Results are in code units.
\textbf{a)} Plane defined by \cref{eq:transition_fast}. %Linearity observed only in the limit of a diffusion dominated growth in the contact region ($\bar{\nu}\gg1$).
\textbf{b)} Generalized formulation, \cref{eq:transition_slow}.
The dashed line is a reference line of slope one passing through the origin.
Results are in normalized units.
\label{fig:morph}
}
\includegraphics[width=\linewidth]{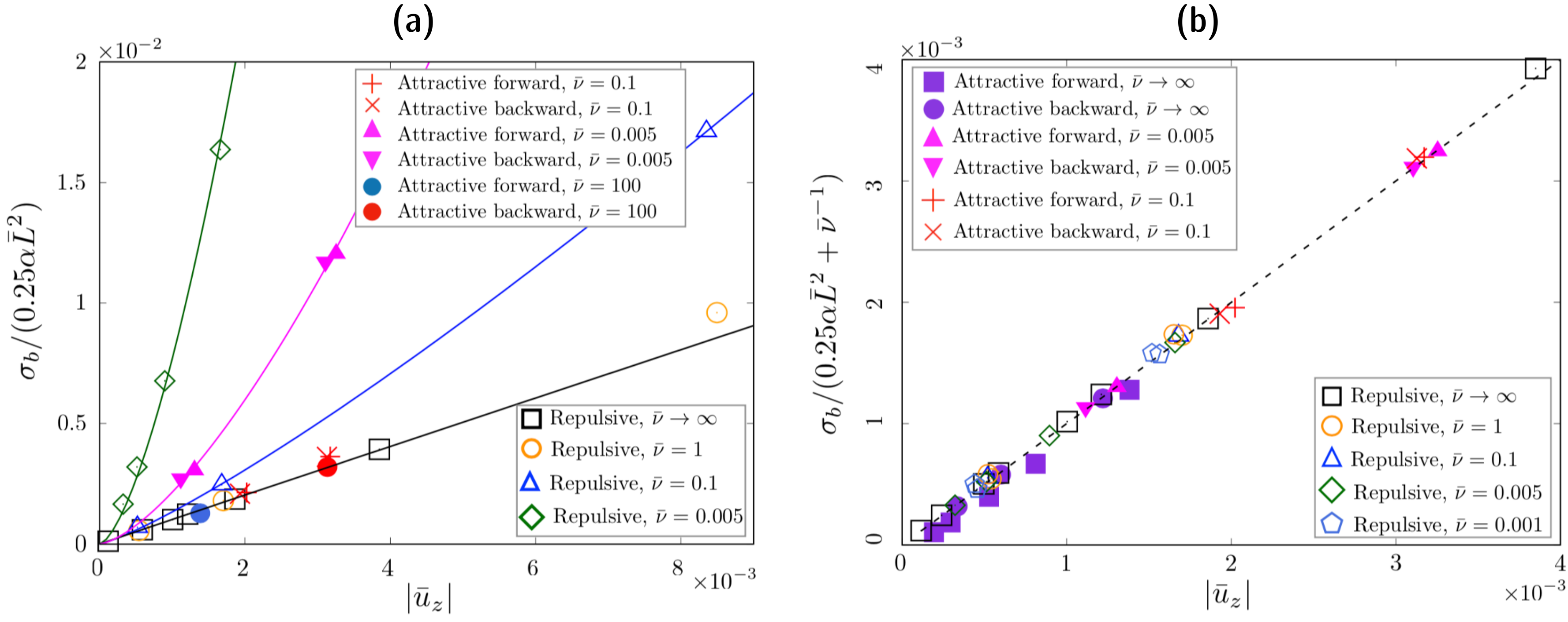}
\end{figure*}

\noindent In this section, we reproduce the derivation of the criterion of \cref{eq:transition_fast}
for cavity formation reported in Ref.~\cite{Kohler2018}, 
and include in addition the effect of surface kinetics.

We start with a local mass balance in the film. 
Neglecting the advection of the concentration and considering a typical film width $h$, one has
 \begin{equation}
 2\pi rh J_d(r) =\pi r^2 J_k\, , 
 \end{equation}
 where $J_d(r) = D\partial_r c$ is the diffusion flux in the liquid film 
and $J_k=|u_z|/\Omega$ is the mass flux entering the crystal. 
From the above relation, we find the concentration profile in the film
\begin{equation}
c(r) = c_b-\frac{|u_z|}{4hD\Omega}(L^2-r^2)\, ,
\end{equation}
where $c(L) = c_b$ is the concentration at the edge of the contact radius.
In particular we obtain an expression for the supersaturation in the center of the contact
\begin{equation}
\label{eq:sigma_center}
\sigma(0) = \sigma_b-\frac{|u_{z}|L^2}{4hDc_0\Omega}\, .
\end{equation}
Let us define the thickness in the center of the contact $\zeta_0=\zeta(r=0)$.
The presence of a cavity implies $U'(\zeta_0)\approx 0$ because 
the depth $\zeta_0$ of the cavity should increase beyond the range of the
disjoining pressure and $\kappa(r=0)<0$ since the cavity is concave.
As a consequence, we expect from \cref{eq:mu_main} that $\Delta\mu(r=0)<0$
in the presence of a cavity.
However, before the formation of the cavity, 
the interface is approximately flat $\kappa(r=0)\approx 0$ and $U'(\zeta_0)> 0$
at least for repulsive interactions, leading to $\Delta\mu(r=0)>0$.
As a consequence, we use the condition of change of sign
of the chemical potential as an estimate of the threshold
for the formation of a cavity, $\Delta\mu^{cav}(r=0) = 0$.

From \cref{eq:concentration,eq:ceq,eq:mu_main}, we express the interfacial chemical potential as 
\begin{equation}
\label{eq:mu_slow_kin}
\Delta \mu(r)=k_BT \Bigr ( -\frac{\mathrm{v}_{z}(r)}{\nu\Omega c_0}+\sigma(r)\Bigl)\, .
\end{equation}
In steady-state, where $\mathrm{v}_z=-u_z$ (\cref{eq:def_vz}),
therefore the condition $\Delta\mu^{cav}(0) = 0$ provides a relation 
between the rigid body velocity of the crystal and the supersaturation in the center:
\begin{equation}
\label{eq:uz_morph}
u_z^{cav} = -\sigma(0)\nu\Omega c_0\, .
\end{equation}
Combining \cref{eq:uz_morph} with the supersaturation profile deduced 
from mass conservation \cref{eq:sigma_center}, we obtain a generalized transition criterion:
\begin{equation}
\label{eq:transition_slow}
|u_{z}^{cav}| = \frac{\Omega c_0 \sigma_b^{cav}}{\alpha L_{cav}^2/(4hD)+\nu^{-1}}\, .
\end{equation}
We have included the phenomenological constant $\alpha\approx 0.6$ to obtain quantitative agreement between this relation and the observations in simulations with fast surface kinetics~\cite{Kohler2018,Gagliardi2018a}.

\subsection{Numerical evaluation of the location of the transition}

\noindent As a first remark, simulations indicate that surface kinetics does not alter the continuous or discontinuous nature of the transition. 
Indeed, the cavity still appears smoothly and continuously in the repulsive case, \cref{eq:potential_Felix}, 
whilst a discontinuous transition with hysteresis is observed
when considering the attractive interaction, \cref{eq:potential_sub}.
We thus apply similar procedures as those described in Refs.~\cite{Kohler2018,Gagliardi2018a} 
to characterize the transition.

We start by plotting the evolution of
the steady-state film width at the center of the contact $\zeta_0$ 
(see, e.g., \cref{fig:cavity}) 
as a function of 
the supersaturation at the boundary of the simulation box  $\sigma_{bc}$. 
In the case of a repulsive potential, the transition is continuous, i.e.,
$\zeta_0$ varies smoothly and does not exhibit any jump
when varying the supersaturation.
The critical threshold value $\sigma_{bc}^{cav}$ is 
then estimated from the intersection of two linear fits: 
one corresponding to the flat growth regime below the transition, where $\zeta_0$ is weakly affected by $\sigma_{bc}$,
and another one above the transition, where $\zeta_0$ increases linearly with $\sigma_{bc}$.
However, in the case of an attractive potential, the transition is discontinuous and
exhibits hysteresis~\cite{Gagliardi2018a}.
In this case, the critical supersaturation $\sigma_{bc}^{cav}$ is obtained from the direct observation of a sharp jump in the value of $\zeta_0$ when varying the supersaturation $\sigma_{bc}$.
Since the transition shows hysteresis as in the diffusion-limited case,
the \enquote{forward} transition  obtained when increasing the supersaturation 
of an initial flat profile, and \enquote{backward} transition observed by decreasing $\sigma_{bc}$ when starting with a profile already presenting a cavity are different.

Once the critical supersaturation $\sigma_{bc}^{cav}$ at the boundary of the simulation box  is obtained, a simulation at $\sigma_{bc}^{cav}$ is performed. 
Then, the critical contact radius $L^{cav}$ is calculated 
by the method discussed in \cref{sec:numerical_methods,app:contact_radius}.
The corresponding supersaturation $\sigma_b^{cav}=\sigma(L^{cav})$ is finally obtained using \cref{eq:concentration}.

%%%%%%%%%%%%%%%%%%%%%%%%%%%%%%%%%%%%%%%%%%%%%%%%%%%%%%%%%%%%%%%%%%%%%%%%%%
%%%%%%%%%%%%%%%%%%%%%%%%%%%%%%%%%%%%%%%%%%%%%%%%%%%%%%%%%%%%%%%%%%%%%%%%%%

\subsection{Data collapse with surface kinetics}
 
\begin{figure*}
\center
\caption{Trajectories within the generalized morphology diagram.
Blue circles and orange triangles: data at the threshold of cavity formation. The dashed line is a reference line of slope one passing through the origin.
Joined red rhombus and magenta pentagons: trajectory of a crystal undergoing a transition.
\textbf{a)}: $\bar{\nu} = 100$.
\textbf{b)}: $\bar{\nu}=0.005$.
Results are in normalized units.
\label{fig:transition}
}
\includegraphics[width=\linewidth]{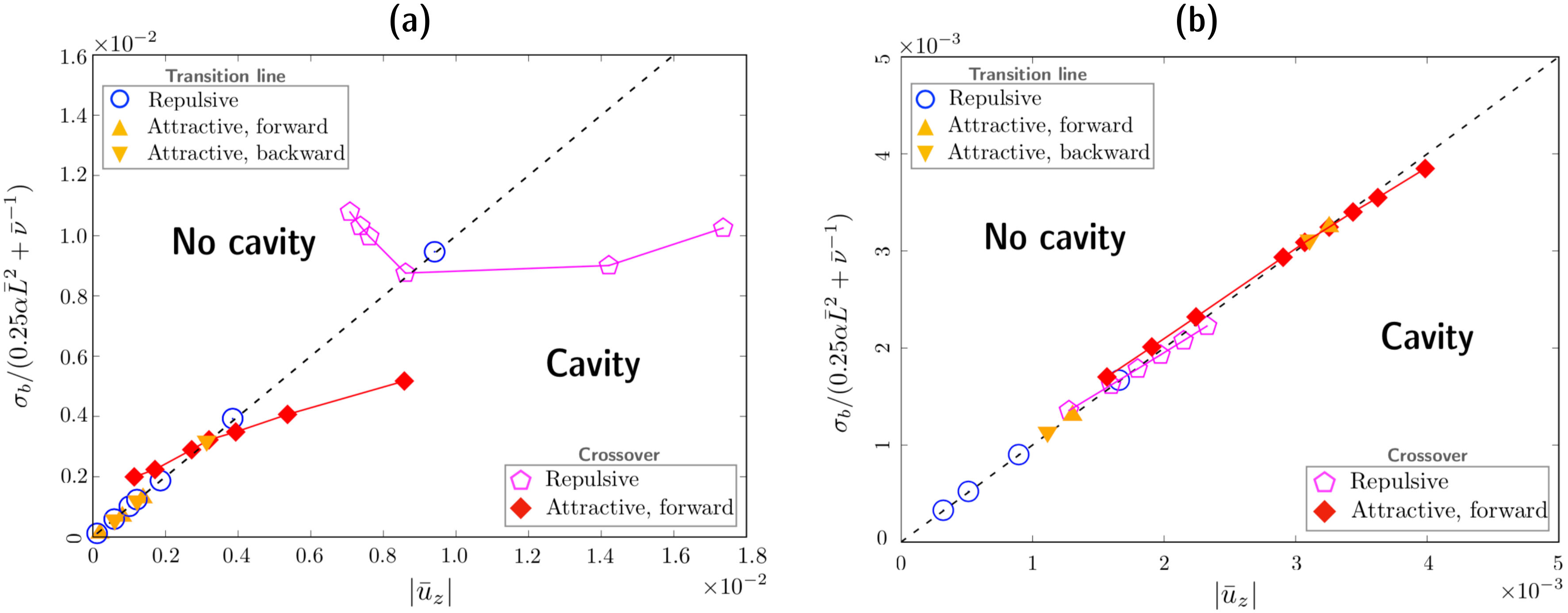}
\end{figure*}

\noindent In \cref{fig:morph} we show simulation results where $L^{cav}$ and  $\sigma_b^{cav}$ are extracted by the procedure discussed above.
The data are reported in normalized units for a large span of values 
of the surface kinetic constant, from $\bar{\nu} = 10^{-3}$ to $\bar{\nu} = 10^2$.  
For the attractive case, we distinguish the results for the two different 
branches of the hysteresis loop (forward and backward).
We also report previous results in the fast attachment limit 
($\bar{\nu}\rightarrow \infty$) from Refs.~\cite{Kohler2018,Gagliardi2018a}.

In \cref{fig:morph}a, we show the 
transition points in the plane defined by the left-hand side and right-hand side of \cref{eq:transition_fast}. 
This representation, denoted as a morphology diagram 
and previously introduced by Ref.~\cite{Kohler2018,Gagliardi2018a},
is expected to lead to data collapse in the diffusion-limited regime. 
Indeed, the simulation results collapse on the same line
in the limit of fast attachment kinetics. 
However, there is clearly no collapse when varying the 
surface kinetic coefficient $\bar\nu$.
Furthermore, the location of the transition depends 
on the functional form of the interaction.
In contrast, all transition points  are shown to collapse 
on the same line in \cref{fig:morph}b when using the
plane coordinates defined by the left-hand-side and the right-hand-side
of \cref{eq:transition_slow}. As expected, this line is straight,
 passes through the origin, and its slope is one.

When varying physical parameters, we expect that
the system will be located in the zone above the 
transition line when there is no cavity, and in the zone below the transition
line in the presence of a cavity. In \cref{fig:transition},
we have reported the trajectory of simulations
with a fixed simulation box $R$ when the supersaturation $\sigma_{bc}$
at the edge of the box is varied. For each point, the contact
size $L$ and the supersaturation at the edge of the contact $\sigma_b$
are measured.
While the transition line is clearly crossed
in the diffusion-limited regime ($\bar\nu\gg 1$) as seen in \cref{fig:transition}a,
the trajectory of the system in the regime limited by surface kinetics ($\bar\nu\ll 1$, \cref{fig:transition}b) is actually along the transition line.

This result can be traced back to the fact that the growth rate
is constant in the surface-kinetics dominated regime,
$|u_z|\approx \nu\Omega c_0\sigma$, with $\sigma\approx\sigma_b$.
This relation is independent of the morphology and is therefore
valid on both sides of the transition point.

In addition, the striking similarities between
\cref{eq:transition_slow,eq:velocity_slowKin}
suggests that the departure from the transition line
in the flat regime could be mainly controlled by finite size effects or by viscosity effects, respectively  via the terms proportional to  $\Delta\mu_{eq}\sim L^{-1}$ 
and $\bar\eta$. 
As a consequence, even in the diffusion-limited regime, the departure from the
transition line on the upper side, for a flat contact, should be small for small viscosities and for large crystals. 
However, when a cavity appears, \cref{eq:velocity_slowKin} ceases to be valid, and 
arbitrary departures from the transition line are possible below the transition line in the diffusion-limited regime.

As a summary, the departure from the transition line is small in the surface-kinetics limited regime. 
In the diffusion-limited regime, departure from the transition line from above (flat contact) are restricted to small crystals or large viscosities.
However, deviations from below, i.e.~in the presence of a cavity, are possible.

%%%%%%%%%%%%%%%%%%%%%%%%%%%%%%%%%%%%%%%%%%%%%%%%%%%%%%%%%%%%%%%%%%%%%%%%%%

\subsection{Transition in the $(L,\sigma)$ plane}

\begin{figure*}
\center
\caption{
Critical supersaturation as a function of contact size.
\textbf{a)} Attractive interaction, \cref{eq:potential_sub}.
The solid lines are analytical predictions using \cref{eq:critical_sup} with $\bar{U}'_{cav} = 5.6\times 10^{-3}$ inferred from data at large contact sizes in the forward transition and for a diffusion limited growth~\cite{Gagliardi2018a}.
Note that for small $\bar{L}$ forward and backward transitions are indistinguishable.\\
\textbf{b)} Repulsive interaction, \cref{eq:potential_Felix}.
The dashed lines are fits with a power law $f(\bar{L}) = c*\bar{L}^{-\beta}$: 
black, $\beta\sim 2.0$;
red, $\beta\sim 3.5$;
blue,  $\beta\sim 4.2$.
\label{fig:critical_sup}
}
\includegraphics[width=\linewidth]{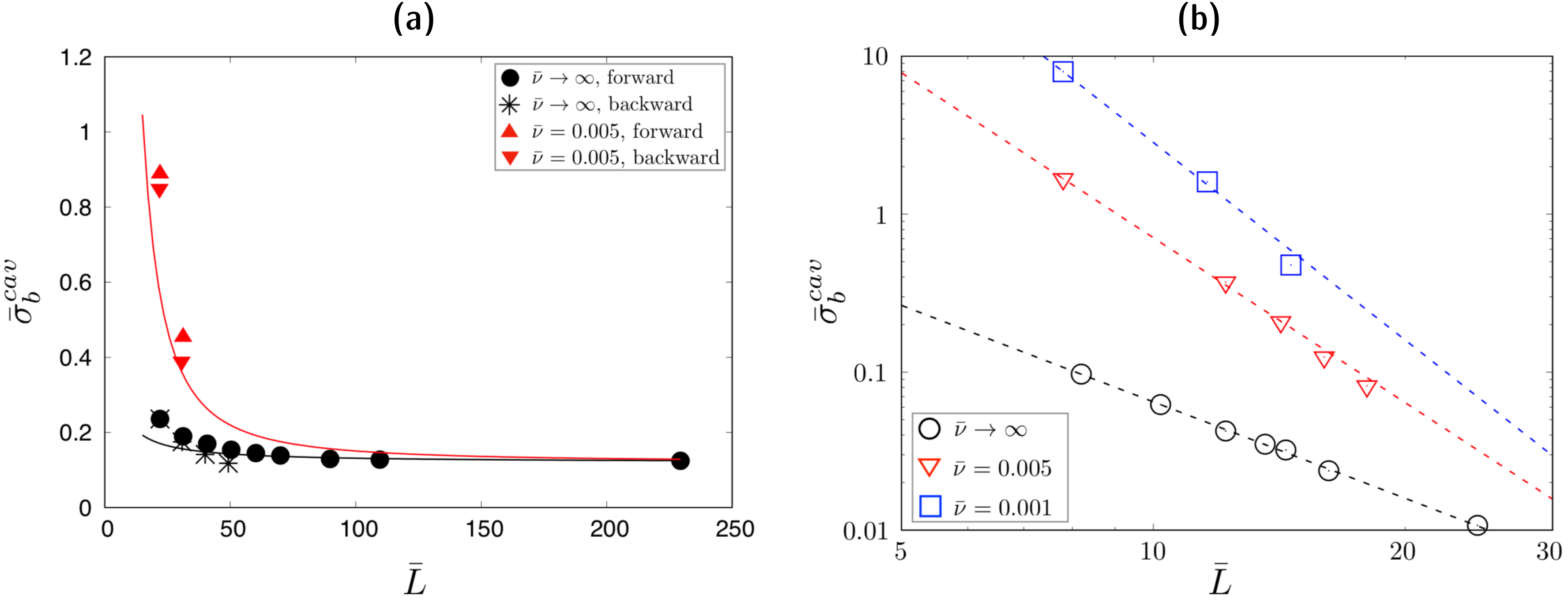}
\end{figure*}

\noindent In growth experiments, such as those of Ref.\cite{Kohler2018}, the 
control parameters are expected to be the contact size $L$ and 
the supersaturation $\sigma_b$. In \cref{fig:critical_sup},
we have reported the transition points in the $(L,\sigma_b)$ plane 
for attractive and repulsive interactions
and for various values of the kinetic coefficient $\bar\nu$.

When the interaction is attractive, the perturbative analysis presented in \cref{sec:lifting,app:perturantion_analysis}
indicates that the crystal exhibit a concave parabolic profile,
associated with an increase of the film width $\zeta_0$ at the center of the contact
before the formation of the cavity. 
Following the same lines as in Ref.\cite{Gagliardi2018},
the onset of cavity formation can be associated with the condition that $\zeta_0$ becomes
large enough for the crystal profile at the center to reach the
regime of spinodal instability where $U''(\zeta_0)<0$.
The instability criterion is therefore  $\zeta_0=\zeta^{cav}$, with $U''(\zeta^{cav})=0$.
In the limit of vanishing viscosity $\eta\rightarrow 0$, we find
\begin{align}
\label{eq:critical_sup}
\sigma_b^{cav} &= \frac{\Omega}{k_BT} {U}'_{cav}\left(1+4\frac{l_0^2-l_{eq}^2}{L^2}\right) 
\nonumber \\
&+ 2\frac{\Omega}{ L}\left[2\tilde{\gamma} \Delta U(h)\right]^{1/2}\!\!\!\!\! ,
\end{align}
where $l_{eq}=(\tilde\gamma/U''(h))^{1/2}$, and ${U}'_{cav}=U'(\zeta^{cav})$.
The details of the derivation of \cref{eq:critical_sup} are reported in \cref{app:critical_sup}.
Using \cref{eq:potential_sub}, we find ${U}'_{cav}=(A/h^3)\bar{U}'_{cav}$,
with $\bar{U}'_{cav}=9/(2^9\pi)\approx 5.6\times10^{-3}$.
In the limit of large contacts, \cref{eq:critical_sup} indicates that 
$\sigma_b^{cav}(L\rightarrow \infty)=(k_BT/\Omega){U}'_{cav}$.
Using the measured asymptotic value $\sigma_{\infty}$ in simulations with large $L$,
we obtain $\bar{U}'_{cav}=(\Omega/k_BT)\sigma_{\infty}\approx 6.4\times 10^{-3}$, 
in fair agreement with the expected value. 
The prediction obtained from \cref{eq:critical_sup} with the latter value of $\bar{U}'_{cav}$ is shown in \cref{fig:critical_sup}a for two different kinetic coefficients $\bar \nu$.
The agreement between \cref{eq:critical_sup,fig:critical_sup}a suggests that the 
critical supersaturation decays to its asymptotic value as 
$\sigma^{cav}_b(L)-\sigma_{\infty}\sim L^{-1}$
in the diffusion-limited regime ($L\gg l_0$), whilst this difference is $\sim L^{-2}$
for slow surface kinetics ($L\ll l_0$). However, the 
accuracy of our simulations and the uncertainty on the asymptotic value 
does not allow us to check these exponents quantitatively.
Finally, simulation results suggest a qualitatively similar behavior 
for the backward transition, with a different asymptotic value $\sigma_\infty$.

In the case where the interaction is repulsive, no analytical prediction is available.
As shown by \cref{fig:critical_sup}b, simulations suggest important differences 
with the scenario predicted by \cref{eq:critical_sup}. First, 
a vanishing critical supersaturation for cavity formation is obtained
for large crystals $L\rightarrow\infty$. This is consistent
with our previous results~\cite{Kohler2018}, and can be understood
intuitively from the absence of trapping of the interface in a potential well 
when the interaction is repulsive.
In addition, the decay of the critical supersaturation
can be fitted with power-laws (dashed lines in \cref{fig:critical_sup}b). We 
find $\sigma_b^{cav}\sim L^{-2}$ in the diffusion driven regime (black dots) 
and $\sigma_b^{cav}\sim L^{-4}$ for the regime dominated by surface kinetics (red and blue dots).

As a summary, discarding physical prefactors, we find
\begin{equation}
\label{eq:summary}
{\sigma}^{cav}_b - {\sigma}_\infty \sim {L}^{-\beta} \, ,
\end{equation}
where $\beta$ and ${\sigma}_\infty$ are listed in \cref{tab:summary}.

 \section{Conclusions}

\begin{table*}
\center
\caption{Summary of the power law regimes observed for 
the transition line in the $(L,\sigma_b)$ plane depending 
on the kinetic process and type of disjoining pressure. 
Notations are those of \cref{eq:summary}.
The asymptotic supersaturarion at large contact sizes, $\bar{\sigma}_\infty$,
 depends on the branch of the hysteresis plot. The reported results correspond to the forward transition. 
\label{tab:summary}}
\renewcommand{\arraystretch}{1.3}
\begin{tabular}{|c|c|c|}
\hline
\diagbox{Dominating process}{Disjoining pressure} &
Repulsive  &
Attractive  \\
\hline
 Diffusion & $\beta \approx 2$, $\sigma_\infty= 0$
&	$\beta \approx 1$, ${\sigma}_\infty \approx \Omega U'_{cav}/k_BT$ (forward)
\\
\hline
 Surface kinetics  & $\beta \approx 4$, $ \sigma_\infty = 0$
& $\beta\approx 2$, ${\sigma}_\infty \approx \Omega U'_{cav}/k_BT$ (forward)
\\
\hline
\end{tabular}
\end{table*}

\noindent In conclusion, we have studied the influence of surface
kinetics on the growth of a crystal in the vicinity of a flat substrate. 
Surface kinetics affects the growth rate within the contact
region: we found a novel regime for contact sizes smaller than $l_0=(Dh/\nu)^{1/2}$,
where the growth rate is independent of the contact size. 
The experimental observation of such a crossover could 
allow one to gain novel quantitative insights on the value of the kinetic constant.
Ultimately, the total displacement of the crystal bulk due to growth within the contact
is always finite. Furthermore, slow surface kinetics does not prevent the formation of a cavity, and the subsequent formation of a rim.
However, the straightforward generalization of the condition for the 
formation of a cavity obtained in Ref.\cite{Kohler2018} appears
to be uninformative in the limit of slow surface kinetics.
Instead, we formulate this condition in terms of a critical supersaturation above which the cavity forms.
This critical supersaturation
is found to be larger for slower surface kinetics, 
and to decrease as a power-law of the contact size.

%%%%%%%%%%%%%%%%%%%%%%%%%%%%%%%%%%%%%%%%%%%%%%%%%%%%%%%%%%%%%%%%%%%%%%%%%%
%%%%%%%%%%%%%%%%%%%%%%%%%%%%%%%%%%%%%%%%%%%%%%%%%%%%%%%%%%%%%%%%%%%%%%%%%%
%%%%%%%%%%%%%%%%%%%%%%%%%%%%%%%%%%%%%%%%%%%%%%%%%%%%%%%%%%%%%%%%%%%%%%%%%%

\section*{Acknowledgement}
The authors wish to acknowledge funding from the European Union's Horizon 2020 research and innovation program under grant agreement No 642976.

\appendix

%%%%%%%%%%%%%%%%%%%%%%%%%%%%%%%%%%%%%%%%%%%%%%%%%%%%%%%%%%%%%%%%%%%%%%%%%%
%%%%%%%%%%%%%%%%%%%%%%%%%%%%%%%%%%%%%%%%%%%%%%%%%%%%%%%%%%%%%%%%%%%%%%%%%%
%%%%%%%%%%%%%%%%%%%%%%%%%%%%%%%%%%%%%%%%%%%%%%%%%%%%%%%%%%%%%%%%%%%%%%%%%%
\section{Model derivation}
\label{app:model}

\noindent The derivation of the thin film equation in the presence of slow surface kinetics follows the same lines as in Ref.~\cite{Gagliardi2018}. Here, we summarize the main steps of the derivation and point out the technical differences that result from the assumption of slow surface kinetics.

A schematic of the system is presented in \cref{fig:sketch}.
We assume a crystal growing (or dissolving) in a liquid solution and in the vicinity of a substrate.
The surface of the substrate is flat and parallel to the $(x,y)$ plane.  
The substrate is homogeneous, immobile, impermeable and inert.
We consider a rigid crystal -- i.e.~neglect elastic effects --, 
with equal and constant densities in the liquid and the crystal $\rho_L =\rho_C$. 
A \emph{disjoining pressure} $U'(\zeta)$ acts between the crystal surface and the substrate~\cite{Israelachvili2011} where $\zeta(x,y,t)$ is the local thickness of the film  along the $z$ axis. 
The disjoining pressure is the derivative of the interaction potential $U(\zeta)$.
The velocity of the bulk of the crystal along the $z$ direction is $u_z$.
 We neglect lateral  motion of the crystal $\mathbf{u}_{xy} = 0$.
Here and in the following, the subscript $_{xy}$ indicates the projection of 
a vector field along the substrate plane. 

The derivation of the thin film model is based on a small slope expansion (also called the lubrication limit)~\cite{Oron1997}. 
This expansion procedure  exploits a disparity of scale in the liquid film, 
namely that the length scale $\ell$ associated to variations 
of the film thickness and of the concentration in the $(x,y)$ plane 
are much larger than the film thickness $\sim h$.
We thus identify a small parameter $\epsilon = h/\ell$.
Spatial coordinates are rescaled as
$x\sim y \sim \ell\sim  h/\epsilon$, and $z\sim h$.
Furthermore, assuming that the typical liquid velocity
parallel to the substrate is of order one %$\mathbf{u}_{Lxy}\sim u_0$, 
we also consistently choose the pressure $p \sim 1/\epsilon$, and time $t\sim 1/\epsilon$. We also assume $c\sim\mathcal{O}(1)$.
To leading order in the expansion, the pressure $p$ and the solute concentration $c$ 
in the liquid do not depend on $z$~\cite{Oron1997}.

Let us recall the main relations  in the lubrication limit as obtained in Ref.~\cite{Gagliardi2018}.
First, to leading order, the liquid flow is a Poseuille flow in the $(x,y)$ plane.
The total flow between the substrate and the crystal
is proportional to the gradient of pressure $\sim(\zeta^3/12\eta)\nabla_{xy} p$,
where  $\eta$ is the viscosity of the liquid.
Then, as a consequence of the conservation of the total mass, 
the rigid motion of the crystal $u_z$ is a source term for the liquid flow in the film
\begin{equation}
\label{eq:mass_cons}
u_{z} = -\nabla_{xy}\cdot\left[ \frac{\zeta^3}{12\eta}\nabla_{xy} p\right ]\, .
\end{equation}
Second, the conservation of crystal units reads
\begin{equation}
\label{eq:cryst_cons}
\frac{\mathrm{v}_{z}}{\Omega}
+ \partial_t[\zeta c]
-\nabla_{xy}\cdot\left[\frac{\zeta^3}{12\eta}c\nabla_{xy}p \right]
%+\frac{\mathbf u_{Cxy}}{2}\cdot\nabla_{xy}[c\zeta]
= \nabla_{xy} \cdot [\zeta{D} \nabla_{xy} c]\, ,
\end{equation}
with $\Omega$ the crystal molecular volume and $D$ the diffusion constant (assumed independent of the concentration). 
Finally, the global force balance on the crystal depends on the contact region and reads~\cite{Gagliardi2018}
\begin{equation}
\label{eq:force_balance}
F_z= \iint_{\text{contact}}\mkern-30mu \mathrm{d}A\, (p-p^{ext} +U'(\zeta))\, ,
\end{equation}
where $F_z$ is an external force acting 
on the crystal, and $\mathrm{d}A =\mathrm{d}x\mathrm{d}y $ and $p^{ext}$ is a constant liquid pressure outside the contact zone.

The link between the growth rate and the supersaturation in the liquid
in physical systems depends on the subjacent microscopic structure of the crystal surface~\cite{Saito1996,Sunagawa1999}.
Here, we assume a linear kinetic law, so that the crystallization rate reads
\begin{equation}
\label{eq:kin_law_general}
v_n-\hat{\mathbf{n}}\cdot\mathbf{u} = \Omega\nu (c-c_{eq})\, ,
\end{equation}
where $v_n$ is the normal velocity of the crystal interface, 
$\nu$ a kinetic constant and $c_{eq}$ the local equilibrium concentration at the liquid-crystal interface.
This relation results from the combination of \cref{eq:growth_rate_def,eq:kin_law_general_main} in the main text.
To leading order in $\epsilon$ \cref{eq:kin_law_general} reads~\cite{Gagliardi2018}
\begin{equation}
\label{eq:kin_law}
\mathrm{v}_{z} = \Omega\nu(c-c_{eq})\, ,
\end{equation}
where $\mathrm{v}_{z}=\mathcal{O}(\epsilon)$ is the local growth rate along $z$ (surface velocity in the reference frame of the crystal),
\begin{equation}
\mathrm{v}_{z} = -\partial_t\zeta - u_z\, .
\end{equation}
In Refs.~\cite{Gagliardi2018,Gagliardi2018a,Kohler2018} the kinetic constant was assumed $\nu\sim \mathcal{O}(1)$  in the lubrication expansion. 
Since $c\sim \mathcal{O}(1)$, this led to a fast local equilibration of the concentration: $c=c_{eq}$.
In contrast, we assume the kinetic constant to be small: $\nu = \mathcal{O}(\epsilon)$. 
As a consequence, all terms in \cref{eq:kin_law} are of the same order, and
\begin{equation}
c=c_{eq} +\frac{\mathrm{v}_{z}}{\Omega\nu} ,
\end{equation}
which is identical to \cref{eq:concentration} in the main text. 
The assumtion $\nu = \mathcal{O}(\epsilon)$, and the resulting \cref{eq:concentration} are the main difference
between the present study and Refs.~\cite{Gagliardi2018,Gagliardi2018a,Kohler2018}.

Finally, following the same lines as in Ref.~\cite{Gagliardi2018},
we consider the dilute limit $\Omega c\ll 1$ in axisymmetric
geometries. This leads to the equations presented in the main text in \cref{sec:model_equations}.

%%%%%%%%%%%%%%%%%%%%%%%%%%%%%%%%%%%%%%%%%%%%%%%%%%%%%%%%%%%%%%%%%%%%%%%%%%
%%%%%%%%%%%%%%%%%%%%%%%%%%%%%%%%%%%%%%%%%%%%%%%%%%%%%%%%%%%%%%%%%%%%%%%%%%
%%%%%%%%%%%%%%%%%%%%%%%%%%%%%%%%%%%%%%%%%%%%%%%%%%%%%%%%%%%%%%%%%%%%%%%%%%

\section{Perturbation to equilibrium for an attractive interaction }
\label{app:perturantion_analysis}

%\begin{figure}
%\includegraphics[width=\linewidth]{figures/figure8}
%\caption{Steady state crystal profile (black line) as obtained by a simulation close to the transition compared to the analytic expression from linear perturbation analysis (red dashed line), \cref{eq:pert}. The contact size -- measured from the criterion $\bar{L}=\max_{\bar{r}} [\partial_{\bar{r}}(\partial_{\bar{r}\bar{r}}\bar{\zeta})]$ -- is $\bar{L}=30.6$ (boundary of the analytical curve).
%Simulation parameters are: $\bar{R} = 40$, $\bar{\sigma}_{bc} = 0.45$, and $\bar{\nu} = 5\times 10^{-3}$. The formation of a cavity would be observed at $\bar{\sigma}_{bc} =0. 53$.
%\label{fig:profile_pert}}
%\end{figure}

\noindent Let us consider steady-state solutions of \cref{eq:evolution_vz}. 
Since in steady state $\partial_t \zeta = 0$, we have $\mathrm{v}_{z}= -u_{z}$ where $u_{z}$ is a constant,
and
\begin{equation}
\label{eq:steady}
0 = B\frac{1}{r}\partial_r [r \zeta \partial_r (\tilde{\gamma} \partial_{rr}\zeta +\frac{\tilde{\gamma}}{r}\partial_r \zeta -U'(\zeta))] + u_z \, ,
\end{equation}
with $B = \Omega^2c_0D/(k_BT)$.
This steady-state equation is identical to that of the diffusion-dominated regime
presented in  Ref.~\cite{Gagliardi2018a}. 
Because of this analogy with the diffusion-dominated scenario, 
the perturbation analysis follows the same steps as in Ref.~\cite{Gagliardi2018a},
and in the following
we only briefly recall the main steps of the derivation.

We consider the case of an attractive crystal-substrate interaction 
and consider a crystal below the transition, i.e., without a cavity.
Below the transition, we expect a small departure from equilibrium 
so that we seek solutions of the steady-state profile 
as a small perturbation of the equilibrium solution: $\zeta(r) =\zeta_{eq}(r) +\delta\zeta(r) $.
Given the attractive interaction, we assume the equilibrium profile to be flat in the contact region, $\zeta_{eq}(r) = h$ with $h$ a constant representing the minimum of the interaction, $U'(h) = 0$.

The equilibrium solution is a particular solution of \cref{eq:steady} with $u_z=0$ which obeys
\begin{equation}
\frac{\Delta\mu_{eq}}{\Omega}= \tilde{\gamma} \partial_{rr}\zeta_{eq} +\tilde{\gamma}/r\partial_r \zeta_{eq} -U'(\zeta_{eq})\, , 
\end{equation}
where $\Delta\mu_{eq}$ is the constant equilibrium chemical potential.
Integrating the above relation and using the fact that $\zeta_{eq}\approx h$
in the center of the contact, the equilibrium chemical potential is given by~\cite{Gagliardi2018a}
\begin{equation}
\label{eq:chemical_pot_eq}
\Delta \mu_{eq}  \approx \frac{2\Omega}{L}\sqrt{2\tilde{\gamma}\Delta U}\, ,
\end{equation}
with $\Delta U = U_\infty - U(h)$ and $U_\infty\approx 0$ is
the constant value of the interaction potential far from the substrate.
By convention, we use $U_\infty=0$.
As a remark, this equation can be re-written as $\Delta \mu_{eq}  \approx \theta^{eq}_{tl}{2\Omega}/{L}$,
where $\theta^{eq}_{tl}\approx\sqrt{2\tilde{\gamma}\Delta U}$  is the equilibrium contact angle.

We now proceed by expanding \cref{eq:steady} to linear order in $\delta\zeta=\zeta(r)-h$ for $u_z\neq 0$. 
Integrating and using \cref{eq:chemical_pot_eq}, we find:
\begin{equation}
\label{eq:pert}
\delta\zeta  =\frac{u_z}{4BhU''(h)}(r^2-L^2 + \frac{4\tilde{\gamma}}{U''(h)}) - \frac{\Delta \mu_b-\Delta\mu_{eq}}{\Omega U''(h)}\, ,
\end{equation}
where we have defined the chemical potential at the edge of the contact radius $L$ as
$\Delta\mu_b = \Delta \mu(L)$.
This expression shows satisfactory agreement with numerical results as  illustrated in \cref{fig:vel_nu}c.%\cref{fig:profile_pert}.

\subsection{Lifting velocity before the transition}
\label{app:perturantion_analysis_vel}

\noindent Expanding the force balance equation \cref{eq:uz}
to linear order in $\delta\zeta$, and using \cref{eq:pert}, we obtain
\begin{equation}
\label{eq:vel_visc_complete}
u_z = \frac{-4Bh(\Delta\mu_{b}-\Delta\mu_{eq})}{(\frac{6B}{h^2}\eta + \frac{1}{2} -\frac{4\tilde{\gamma}}{L^2U''(h)})L^2\Omega}\, .
\end{equation}
From the relation between chemical potential, supersaturation and local growth rate, \cref{eq:concentration,eq:ceq,eq:mu_main}, together with the steady state condition $\mathrm{v}_{z} = -u_{z}$, we have:
\begin{equation}
\label{eq:mu_slow}
\Delta\mu(r=L)=\Delta\mu_b = k_BT(\sigma_b +\frac{u_{z}}{\Omega c_0\nu})\, .
\end{equation}
Inserting this expression in \cref{eq:vel_visc_complete} and neglecting the term of order $1/L^2$, we finally obtain \cref{eq:velocity_slowKin} in the main text. 
Note that, as already observed in the derivation of the generalized morphology diagram (\cref{sec:morph_D_derivation}), 
the relation between $\Delta \mu$ and $\sigma$, \cref{eq:mu_slow}, 
is the sole difference between \cref{eq:vel_visc_complete} 
and the expression reported in Ref.~\cite{Gagliardi2018a}.

\subsection{Critical supersaturation}
\label{app:critical_sup}

\noindent Following the same lines as in Ref.\cite{Gagliardi2018a}, 
we consider that a cavity appears when $\zeta_0>\zeta^{cav}$ 
where $\zeta_0 = \zeta(r=0)$, and $\zeta^{cav}$ is the thickness above which 
a flat film undergoes a spinodal instability.  
By definition of the spinodal instability, $\zeta^{cav}$ corresponds to
the inflection point of the potential $U''(\zeta^{cav}) = 0$.

Computing \cref{eq:pert} in $r=0$ we have
\begin{equation}
\label{eq:critical_mu}
\Delta\mu_b^{cav}-\Delta\mu_{cav} = \Omega U'_{cav}\left[\frac{(12\bar{\eta} + 1)L^2-\frac{8\tilde{\gamma}}{U''(h)}}{(1-12\bar{\eta})L^2}\right]\, ,
\end{equation}
where $U'_{cav} = \delta\zeta^{cav}U''(h) $.
The latter expression of $U'_{cav}$ is the linear term of an expansion
of the disjoining pressure $U'(\zeta^{cav})$ for small $\delta\zeta^{cav}$. 
Instead of using this linear term only, we choose to use the full nonlinear expression of the disjoining pressure $U'_{cav} = U'(\zeta^{cav})$. 
This minor reformulation -- which was not performed in Ref.~\cite{Gagliardi2018a} --
allows one to reach better quantitative accuracy, as discussed in the main text.

Using \cref{eq:mu_slow,eq:vel_visc_complete}, we finally rewrite \cref{eq:critical_mu} as:
\begin{equation}
k_BT\sigma_b^{cav} -\Delta\mu_{eq} = \Omega U'_{cav}\frac{12\bar{\eta} + 1  + 8\frac{l_0^2}{L^2} -\frac{8\tilde{\gamma}}{U''(h)L^2}}{1+12\bar{\eta}}\, ,
\end{equation}
where $l_0=(Dh/\nu)^{1/2}$ was introduced in the main text.
In the limit where the viscosity is small 
and using \cref{eq:chemical_pot_eq} to express the equilibrium chemical potential, 
we obtain \cref{eq:critical_sup} in the main text.

%%%%%%%%%%%%%%%%%%%%%%%%%%%%%%%%%%%%%%%%%%%%%%%%%%%%%%%%%%%%%%%%%%%%%%%%%%
%%%%%%%%%%%%%%%%%%%%%%%%%%%%%%%%%%%%%%%%%%%%%%%%%%%%%%%%%%%%%%%%%%%%%%%%%%
%%%%%%%%%%%%%%%%%%%%%%%%%%%%%%%%%%%%%%%%%%%%%%%%%%%%%%%%%%%%%%%%%%%%%%%%%%

\section{Lifting dynamics of a crystal on a substrate in the absence of gravity}
\label{app:finite_lifting}

\noindent In \cref{sec:lifting}, we obtained a relation for the evolution of the height $H(t)$ of the center of an isotropic crystal growing in the vicinity of a flat substrate:
\begin{equation}
\label{eq:vel_h}
\frac{d H}{dt} = \frac{u_0l_0^2}{(\frac{3}{2}\bar{\eta}+\frac{1}{8})\left(R(t)^2-H(t)^2\right) + l_0^2}
\end{equation}
with $u_0 = \nu\Omega c_0 \sigma$ the velocity of the free surface away from the contact region in the macroscopic limit (vanishing curvature) and 
\begin{equation}
\label{eq:r_evol-isotropic}
R(t) = R_0 + u_0 t\, ,
\end{equation}
is the radius of the crystal. 
The initial conditions are $R(t=0) = R_0$ and $H(t=0) = H_0$.  
A schematic of the system is shown in \cref{fig:sketchCS}.

Using \cref{eq:r_evol-isotropic}, 
we may rewrite \cref{eq:vel_h} as
\begin{equation}
\label{eq:lifting_intermediate}
\frac{dH}{dR} = \frac{l_0^2}{(\frac{3}{2}\bar{\eta} + \frac{1}{8})(R^2-h^2) + l_0^2}\, ,
\end{equation}
We then perform a change of variable 
\begin{subequations}
\begin{align}
\mathcal{H} &= (\frac{3}{2} \bar{\eta}+ \frac{1}{8})^{1/2}\frac{H}{l_0}\, ,\\
 \mathcal{R}&= (\frac{3}{2}\bar{\eta} + \frac{1}{8})^{1/2}\frac{R}{l_0}\, ,
\end{align}
\end{subequations}
to write \cref{eq:lifting_intermediate} in a normalized form
\begin{equation}
\frac{d\mathcal{R}}{\mathcal{H}} = \mathcal{R}^2 - H^2 + 1\, .
\end{equation}
Using the condition  $R\rightarrow\infty$ when $t\rightarrow\infty$, 
and defining $\mathcal{H}_\infty$ as the asymptotic value of $\mathcal{H}$ at long times, 
the solution of the above equation reads
\begin{equation}
\label{eq:R_lifting}
\mathcal{R} = \mathcal{H} + \frac{2}{\sqrt{\pi}}\frac{e^{\mathcal{H}^2}}{\mathrm{Erfi}[\mathcal{H}_\infty]- \mathrm{Erfi}[\mathcal{H}]}\, .
\end{equation}
We used the imaginary error function defined as
\begin{equation}
\mathrm{Erfi}[z] = \frac{\mathrm{Erf}[iz]}{i} = \frac{2}{\sqrt{\pi}}\int_{0}^z e^{-(is)^2}\, \mathrm{d}s\, ,
\end{equation}
which obeys
\begin{align}
\lim_{z\rightarrow 0}\mathrm{Erfi}[z] &= 0\, ,\label{eq:limit_erfi}\\
\lim_{z\rightarrow \infty}\mathrm{Erfi}[z] &= \frac{e^{z^2}}{\sqrt{\pi} z}\, .
\end{align} 
Since at $t = 0$, we have $\mathcal{R} = \mathcal{R}_0$ and $\mathcal{H} = \mathcal{H}_0$ with $\mathcal{H}_0>0$, \cref{eq:R_lifting} at $t=0$ reads
\begin{equation}
\label{eq:main_lifting}
\mathcal{H}_\infty = \mathrm{Erfi}^{-1}\left[\mathrm{Erfi}[\mathcal{H}_0] + \frac{2}{\sqrt{\pi}}\frac{e^{\mathcal{H}_0^2}}{\mathcal{R}_0 -\mathcal{H}_0}\right]\, .
\end{equation}

Let us now reformulate the result in terms of the initial contact angle.
We denote the initial angle between the height and the radius 
of the crystal  (see \cref{fig:sketchCS}) as $\theta(t = 0) = \theta^{eq}$. 
In the special case of growth after heterogeneous
nucleation, this angle is the equilibrium one, $\theta^{eq} = \pi - \theta_{tl}^{eq}$, 
with $\theta_{tl}^{eq}$ the equilibrium contact angle.
We then have
\begin{equation}
\mathcal{R}_0 = -\frac{\mathcal{H}_0}{\cos(\theta^{eq}_{tl})}\, .
\end{equation}
Assuming in addition that $H_0\ll l_0$ (i.e.~$\mathcal{H}_0\ll 1$), we obtain
\begin{equation}
\label{eq:lifting_final}
\mathcal{H}_\infty \approx \mathrm{Erfi}^{-1}\left[\frac{-2\cos(\theta^{eq}_{tl})}{\sqrt{\pi}\mathcal{H}_0(1 +\cos(\theta^{eq}_{tl}))}\right]\, .
\end{equation}

%%%%%%%%%%%%%%%%%%%%%%%%%%%%%%%%%%%%%%%%%%%%%%%%%%%%%%%%%%%%%%%%%%%%%%%%%%
%%%%%%%%%%%%%%%%%%%%%%%%%%%%%%%%%%%%%%%%%%%%%%%%%%%%%%%%%%%%%%%%%%%%%%%%%%
%%%%%%%%%%%%%%%%%%%%%%%%%%%%%%%%%%%%%%%%%%%%%%%%%%%%%%%%%%%%%%%%%%%%%%%%%%

\section{Contact radius determination}
\label{app:contact_radius}

\begin{figure}[ht!]
\center
\caption{Top panel: 1D and 2D curvatures as a function of the radius for a repulsive (red and orange) and an attractive (blue and light-blue) interaction.  Bottom panel: corresponding steady-state profiles (below the transition) along the radial coordinate $r$. The attractive case is produced considering a forward transition (the profile is flat at the beginning of the integration). 
Attractive potential: $\bar{\sigma}_{bc} = 0.15$ ($\bar{\sigma}_{bc}^{cav}\approx 0.34$).
Repulsive potential: $\bar{\sigma}_{bc} =0.105$ ($\bar{\sigma}_{bc}^{cav}\approx 0.13$).
Vertical lines indicate the position of the maximum of the derivative of the 1D curvature, 
$\bar{L}=\max_{\bar{r}} [\partial_{\bar{r}}(\partial_{\bar{r}\bar{r}}\bar{\zeta})]$. 
For the attractive potential $\bar{L} = 18.4$, for the repulsive potential $\bar{L} = 10.8$. 
Size of the simulation box $\bar{R} = 30$, kinetic constant $\bar{\nu} =100$.
\label{fig:curv}}
\includegraphics[width=\linewidth]{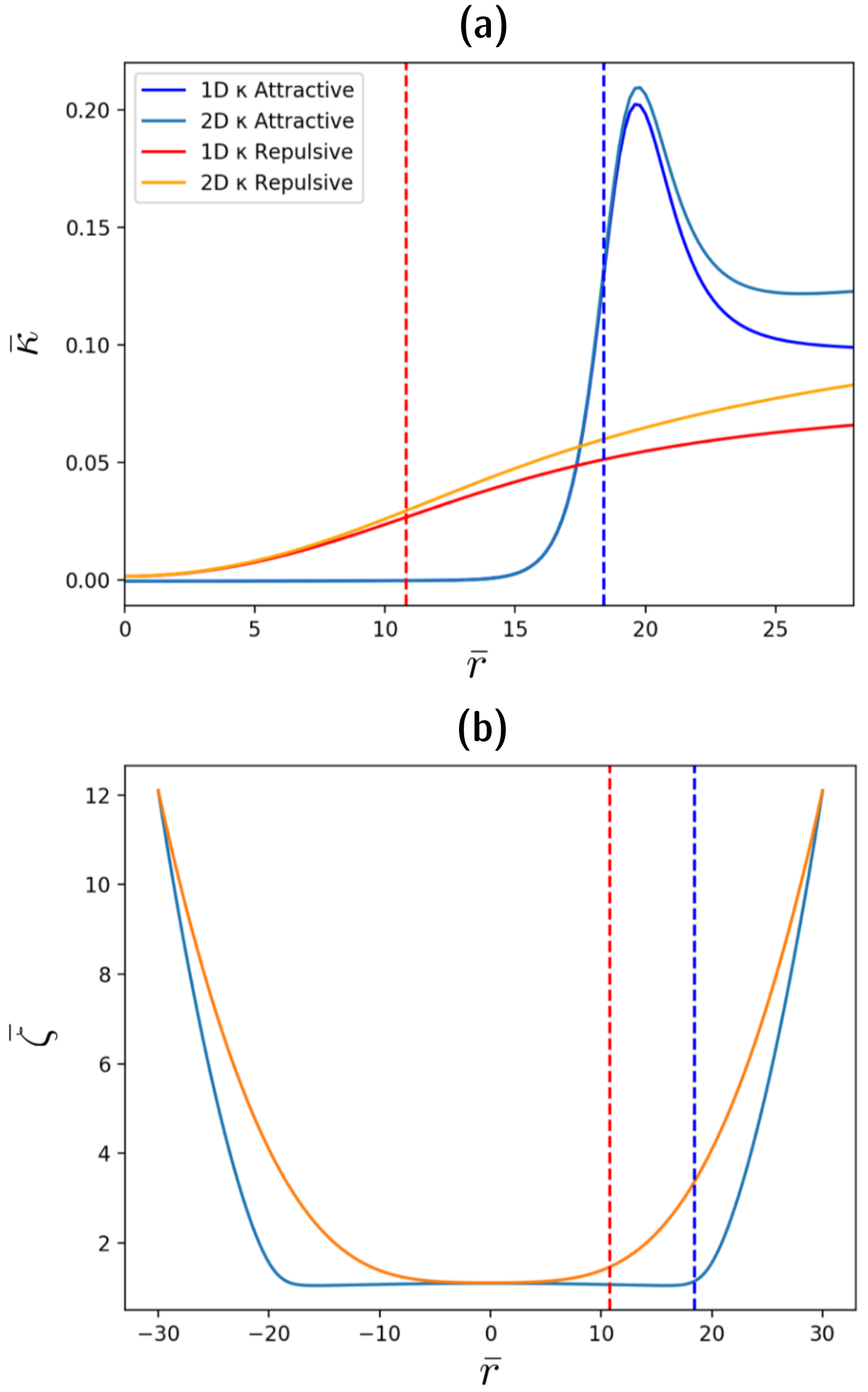}
\end{figure}

\noindent In our previous studies~\cite{Kohler2018,Gagliardi2018a},
the contact radius, $L$,  was determined at the transition point
using \emph{ad hoc} definitions which depend on the type of interaction considered.
We chose  $\zeta(L) = h+20\bar{\lambda}h$ for the repulsive interaction \cref{eq:potential_Felix}, 
and $\zeta(L) = h + 0.001$ for the attractive case \cref{eq:potential_sub}~\cite{Kohler2018,Gagliardi2018a}. 
These definitions proved to provide results that are robust 
with respect to variations of the thickness $\zeta_{bc}$ of the liquid film at the boundary of the simulation box.
However, we would like a more intuitive definition, which would be valid
both for the repulsive and the attractive cases.
We therefore propose that the contact radius is given by the position 
of the maximum of the derivative of the 1D curvature of the crystal profile, 
$\kappa_{1D} = \partial_{rr}\zeta$, $L = \max_{r}[\partial_r \kappa_{1D}]$.
The position of the edge of the contact is shown in \cref{fig:curv} 
for both attractive and repulsive disjoining pressures. 
In the figure, we also show that the 1D and 2D curvatures
behave in a similar fashion. We therefore chose the simplest
condition based on the 1D curvature.

This definition can be understood intuitively at equilibrium, where the chemical potential $\Delta\mu_{eq}$ is a constant. We then have from \cref{eq:mu_main}
\begin{equation}
 \tilde{\gamma}\kappa = \frac{\Delta\mu_{eq}}{\Omega} + U'(\zeta)\, .
\end{equation}
Hence, $\kappa$ is proportional to $U'(\zeta)$ up to an additive constant.
Since the profile is roughly flat in the center, we have $\kappa(r=0)\sim 0$,
for both the repulsive and the attractive cases.

For the purely repulsive interaction, as we move away
from the center of the contact by increasing $r$, and $\zeta$ increases to infinity,
$U'(\zeta)$ increases from negative values to zero.
Thus we expect a monotonous increase of the curvature with the distance from the center,
with a maximum increase located near the edge of the contact where the potential $U$
varies quickly.

In contrast, for the attractive case, we expect an initial increase  of $U'$ when increasing $\zeta$, followed by a long-range decrease  of $U'$ for larger $\zeta$.
As a consequence, we expect the maximum of the derivative of the curvature
to be reached for $\zeta$ in the edge of the contact (and smaller than the value 
for which $U''(\zeta)=0$, which corresponds to a cancellation of 
$\partial_r \kappa_{1D}$ and to the maximum of $\kappa_{1D}$).

Finally, we observe in  \cref{fig:curv} that a qualitatively similar behavior of the curvature is maintained away from equilibrium. 
Therefore, we use the same definition as a signature of the edge of the contact region
in non-equilibrium situations.

\newpage

\bibliographystyle{elsarticle-num-names}
\biboptions{sort&compress}
\bibliography{slowK}






% \end{thebibliography}

\end{document}